\documentclass[reprint,
 amsmath,amssymb,
 aps,
]{revtex4-2}
\usepackage{physics}% for braket julin
\usepackage{graphicx}% Include figure files
\usepackage{xcolor}
\definecolor{purple}{rgb}{0.5,0,0.5}

\graphicspath{ {images/}, {figs/} }
\usepackage{dcolumn}% Align table columns on decimal point
\usepackage{bm}% bold math
\usepackage[mathlines]{lineno}% Enable numbering of text and display math
\usepackage{amssymb}
\usepackage[toc,page]{appendix}
\usepackage{booktabs}
\usepackage{tabularx}
\usepackage{appendix}
\usepackage{xcolor}
\usepackage{array}
\usepackage{comment}

\usepackage[caption=false]{subfig}
\makeatletter

\newcommand{\colorcaption}[2][]{%
  \begingroup%
  \renewcommand{\@caption@fignum@sep}{ (color online). }%
  \caption[#1]{#2}%
  \endgroup%
}
\makeatother

\makeatletter
\renewcommand{\maketag@@@}[1]{\hbox{\m@th\normalsize\normalfont#1}}%
\makeatother

\usepackage{mathtools}
\usepackage{empheq}
\usepackage{refcount}

\usepackage{multirow}

\usepackage{hyperref}% add hypertext capabilities
\hypersetup{
	colorlinks,
	linkcolor={red!50!black},
	citecolor={green!50!black},
	urlcolor={blue!80!black}
}

\usepackage{bm}% bold math
\usepackage{amssymb}
\usepackage{amsmath}
\usepackage{epsfig}
\usepackage{braket}
\usepackage{epstopdf}
\usepackage{graphicx,epsfig}
\usepackage{mathrsfs}
\usepackage{dcolumn}
\usepackage{color}
\usepackage[dvipsnames]{xcolor}
\usepackage{natbib}
\usepackage{CJK}
\usepackage{bbm}
\usepackage{booktabs}
\usepackage{diagbox}
\usepackage{amssymb}
\usepackage{float} 
\usepackage{makecell}
\usepackage{verbatim}
\usepackage{placeins}
\usepackage{ragged2e}
\usepackage{lipsum} % for dummy text

\makeatletter
\long\def\@makefntext#1{%
  \parindent 1em%
  \noindent
  \hb@xt@1.8em{\hss\@makefnmark}#1%
}

\begin{document}
%\preprint{APS/123-QED}

\title{Emergence and transition of incompressible phases in decorated Landau levels}

\author{Bo Peng}
\affiliation{School of Physical and Mathematical Sciences, Nanyang Technological University, 639798 Singapore}

\author{Yuzhu Wang}
\affiliation{School of Physical and Mathematical Sciences, Nanyang Technological University, 639798 Singapore}

\author{Bo Yang}
\email{yang.bo@ntu.edu.sg}
\affiliation{School of Physical and Mathematical Sciences, Nanyang Technological University, 639798 Singapore}

%\date{\today }
\begin{abstract}
A single Landau level (LL) dressed with periodic electrostatic potentials can realize a plethora of interacting topological phases where the Hall conductivity generally does not equal to the LL filling factor. Their physics can be captured by a new family of flat topological bands: decorated Landau levels (dLL) from imposing an electrostatic delta potential lattice within a single LL. With $p/q$ magnetic fluxes per unit cell, there are $q$ dispersive bands and $p-q$ zero energy bands forming the dLL. When the electrostatic potential strength dominates the electron-electron interaction, band mixing is suppressed and the dispersion bands consist of ``localized states" with vanishing total Chern number. Nevertheless these dispersive bands can have highly nontrivial Berry curvature distribution, and even non-zero Chern numbers when $q>1$. Interestingly even in the limit of large short range interaction, band mixing between dLL and dispersion bands can be strongly suppressed at low filling factor, leading to robust topological phases within the dLL stabilized by the one-body potential. The dLL and the associated dispersive bands can serve as minimal theoretical models for correlated physics in lattice or moir\'e  systems; they are also highly tunable experimental platforms for realizing rich phase diagrams of exotic 2D quantum fluids.   

\end{abstract}

\maketitle

\textit{Introduction--} Landau levels (LL) are realized in two-dimensional electron gas systems with a strong perpendicular magnetic field\cite{LL1,LL2,LL3,LL4}. They serve as the simplest examples of flat topological Chern bands with uniform quantum geometry both in real and momentum space\cite{flatband1,flatband2,flatbandMoTe2}. Recent experimental and theoretical progress of Chern insulators with more general Chern bands in lattice or moir\'e systems open up new possibilities of realizing emergent fractional topological phases in a wide variety of quantum materials and experimental platforms\cite{latticeinterplay1,latticeinterplay2,IFB1,IFB2,flatter1,flatter2,topo2,topo3}. The relationship between LL and their lattice analog without an external magnetic field is of both fundamental theoretical interest and experimental relevance\cite{FCI1,FCI2,FCI3,FCI4,FCI5}. It has been proposed that robust fractional quantum Hall phases can be more easily realized in general Chern bands sharing similar characteristics as LL at the single particle level\cite{singleChernband1,singleChernband2}. There is also interesting numerical evidence of distinct features in general Chern bands, where fluctuating Berry curvature leads to gapped phases at unexpected band filling, softening of gapped neutral excitations, as well as short lifetime of long wavelength geometric excitations of fractional phases\cite{flatter2,BerryCurvature1,lifetime1,lifetime2}. 

While in clean LL the magnetic length is the only dominant length scale, rich physics emerges in Chern insulators due to the competing length scales, where the lattice constant can no longer be ignored\cite{length1,length2,length3,length4}. Rather than focusing on how general Chern bands without an external magnetic field can mimic the physical properties of the LL, it is also useful to explore how LL can capture the physics of lattice or moir\'e Chern bands where periodicity plays an important role. Modulating LL with periodic magnetic\cite{periodicmodulation1,periodicmodulation2,periodicmodulation4} or electrostatic fields has the benefits of experimental feasibility and analytic tractability for engineering unconventional topological platforms beyond simple LL. More importantly, it opens up new pathways to understand fundamental differences and similarities between Chern bands with and without the magnetic field\cite{withoutChern1,withoutChern2,withoutChern3,withoutChern4}. 

In this Letter, we systematically analyse the band splitting of a single LL with either periodic or disordered electrostatic potentials, and demonstrate the nontrivial interplay between strong interaction, localization effect, and single particle geometric properties. By tuning the filling factor and the strength of the electrostatic potential, intricate phase diagrams of both compressible and incompressible phases can be realized. Notably for the incompressible topological phases, the Hall conductivity generally differs from the filling factor. Chern bands and topologically trivial bands can be realized, and both have non-uniform Berry curvature distributions similar to those in the lattice systems. The emergent fractional topological phases thus reveal the underlying mechanisms of unconventional gapped phases in some lattice systems\cite{sunkai}. Our general approach also introduces a family of \emph{ideal flat Chern bands}, termed as decorated Landau levels (dLL) with tunable geometric properties, more readily implementable in experiments~\cite{footnote} compared to other approaches such as modulated LLs or $g=2$ quantum Hall systems\cite{g2qh,periodicmodulation1,periodicmodulation2,periodicmodulation4,dllexp1,dllexp2,dllexp3}. We further studied the neutral geometric excitations of the incompressible phase in dLLs, showing the interaction energy scale is significantly smaller compared to that in the conventional LL, and the lifetime of graviton-like excitations could be significantly shorter, analogous to the behavior observed in moir\'e systems~\cite{lifetime2}.

\textit{Single particle Hamiltonian--} We start with a simple model with a set of delta potentials in real space
\begin{equation}
\label{vlocal}
    \hat{V}_{local}=\sum_i\delta (\hat{\textbf{r}}-\textbf{r}_i),
\end{equation}
in which $\textbf{r}_i$ is the position of $i-$th delta potential. We focus on systems with a large magnetic field so that LL mixing can be ignored, and Eq.(\ref{vlocal}) is a one-body Hamiltonian within a single LL.
\begin{figure}[t]
\centering
\includegraphics[width=1.0 \linewidth]{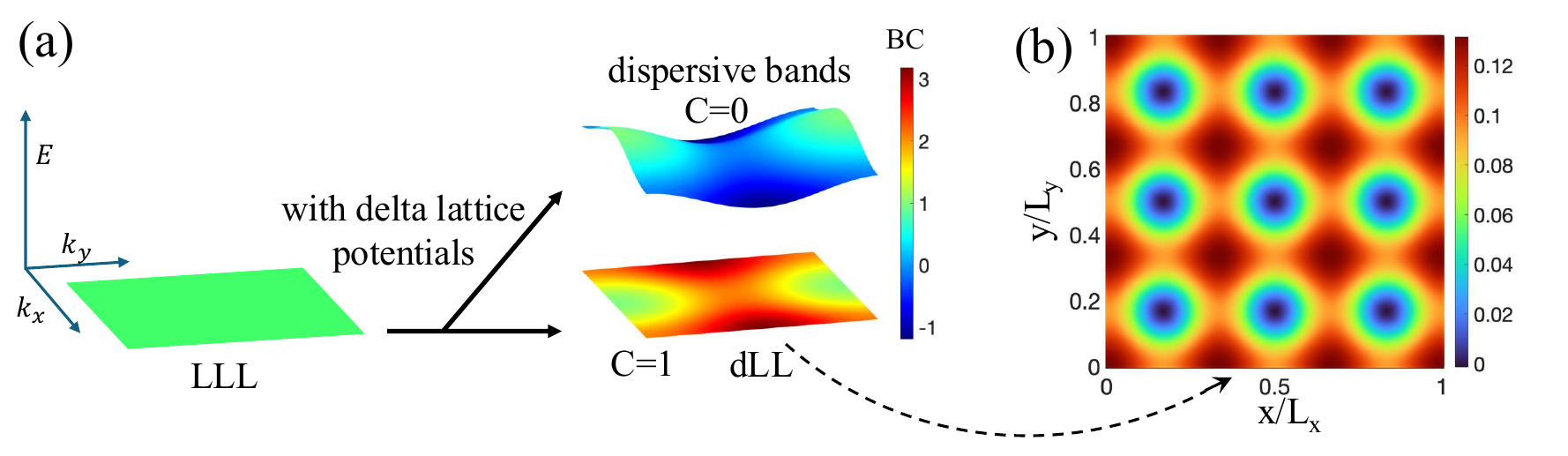}
\caption{(a) Schematic illustration of LLL splitting into dispersive bands and the dLL due to the local potential lattice. BC: Berry Curvature. The color scale gives the Berry Curvature. (b) Real space profile of the dLL.
}
\label{fig1}
\end{figure}
Let the number of delta potentials be $N_{\delta}$ and the number of flux quanta be $N_{\phi}=\frac{B\mathcal A}{2\pi\ell_B^2}$, with $B$ being the uniform magnetic field and $\mathcal A$ the area of the Hall manifold; $\ell_B=\sqrt{\hbar/eB}$ is the magnetic length ($e$ is the electron charge). The case of $N_\delta\ge N_\phi$ has been well studied, showing the system exhibiting robust translational invariance even in the presence of the delta potentials\cite{invariance}. Here we focus on the case of $N_{\delta} < N_{\phi}$, where the Riemann–Roch theorem guarantees the existence of $N_{\phi}-N_{\delta}$ exact zero-energy states~\cite{RiemannRoch,torusHaldane}. The LL thus splits into two subspaces, classified according to whether the single particle energy vanishes. The subspace spanned by these zero-energy states can be regarded as a \textit{dLL} carrying Chern number 1. The other subspace consists of states localized by the delta potentials and carries zero total Chern number~(see fig.~\ref{fig1}).

While most of the results in this work are valid even when the delta potentials are randomly distributed, as we will discuss towards the end, we start with the special cases where a lattice is formed with one delta potential per unit cell. The general case is when each unit cell contains $p/q$ fluxes with co-prime integers $p>q$. In the magnetic field, the electron’s motion is separated into guiding centers $\hat{\textbf{R}}$ and cyclotron motion $\bar{\textbf{R}}$, defined as $\hat{\textbf{R}}^a=\hat{\textbf{r}}^a+\epsilon^{ab}\bm{\pi}_b\ell_B^2$ and $\bar{\textbf{R}}^a=-\epsilon^{ab}\bm{\pi}_b\ell_B^2$, where ${\pi}_a=-i\partial_a-eA_a$ with $A_a$ being the vector potential. We consider an enlarged unit cell containing  $q$ delta potentials in real space. Thus, it also contains $p$ magnetic flux quanta. The corresponding reciprocal vectors $\textbf{b}_{1(2)}$ satisfies $|\textbf{b}_1 \times \textbf{b}_2|=2\pi p \ell_B^{-2}$; a maximal set of commuting magnetic translation operators can be given by $\hat{T}_{\textbf{G}_{mn}}=e^{i \textbf{G}_{mn}\cdot \hat{\textbf{R}}}$, with $\textbf{G}_{mn}=m\textbf{b}_1+n\textbf{b}_2$ where $m$ and $n$ are integers. Clearly $\hat{T}_{\textbf{G}_{mn}}$ also commutes with Eq.(\ref{vlocal}), and their simultaneous eigenstates can be denoted as $|\vec k,i\rangle$, where $\vec k$ is the momentum within the first Brillouin zone defined by $\bm b_{1(2)}$, and $i$ is the band index running from $1$ to $p$ given there are $p$ magnetic fluxes (thus $p$ states within a single LL) per unit cell. The $q$ delta potentials per unit cell lead to $q$ non-zero energy bands carrying a total Chern number of zero, and we term these $q$ bands as the dispersive bands, some of  which may not be trivial, carrying non-zero Chern numbers\cite{supple}. The rest $p-q$ degenerate bands at zero energy form the dLL carrying a total Chern number of unity. With realistic periodic one-body potentials instead of the delta potential lattice, the degeneracy of the $p-q$ bands will be lifted\cite{supple}. For both the Chern bands and the trivial bands, there is a non-trivial Berry curvature distribution within the Brillouin zone. 

%{\color{blue}In the non-interacting limit, only the electrons in the dLL ($p-q$ non-trivial bands) contribute to the Hall conductivity.  When the number of electrons is $N_{e}=N_{\phi}(p-q)/p$ and they fully fill the dLL, an IQHE phase will form with the Hall conductivity $\sigma_{xy}=1$ (unless otherwise mentioned, Hall conductivity is given in unit of $e^2/h$) at the electron filling $\nu=(p-q)/p$ of the LL. In contrast, when $N_{e}= N_{\phi}q/p$ and the electrons fill up $q$ trivial bands, we will have a trivial insulator with zero Hall conductivity. }

\textit{Interaction as a small perturbation--} We now add an interaction \emph{within} the LL with the following Hamiltonian:
\begin{equation}\label{vint}
    H=V_{int}+\lambda V_{local}.
\end{equation}
Let us first consider $|\lambda|\gg 1$ so that the delta potentials are the dominant energy scale. In this limit, the interaction does not mix states between the zero-energy and dispersive bands. For attractive~($\lambda<0$) potentials electrons first fill the $q$ trivial bands; any remaining electrons occupy the topological band, generally giving a non-zero Hall conductivity. Conversely, for repulsive~($\lambda>0$) potentials, the topological band is filled up first.

We define $\nu_l$ as the filling factor of entire LL, given by $\nu_l=N_e/N_{\phi}$, and $\nu_{dl}$ as the filling factor of dLL, given by $\nu_{dl}=N_{e,dl}/\left(N_{\phi}-N_{\delta}\right)$, where $N_{e,dl}$ is the number of electrons in the dLL. When $\lambda<0$ and $N_e < N_{\delta}$, all electrons are confined within the subspace of $q$ dispersive bands. The dispersion of these band dominates the repulsive interaction in this limit. Thus if the fermi surface does not lie within the band gaps (i.e. when $q>1$), a gapless Fermi liquid phase is realized with an anomalous Hall conductance given by the total Berry curvature enclosed by the Fermi surface\cite{Fermiliquid1,Fermiliquid2}. In particular, an insulating phase occurs when $N_e = N_{\delta}$ with a vanishing Hall conductivity. Further increasing the electron number, we fully occupy the trivial bands while the remaining electrons populate the topological band, leading to $\nu_{dl}=\left(N_e-N_{\delta}\right)/\left(N_{\phi}-N_{\delta}\right)$. Notably when $\nu_{dl}=\nu_t$, the common fractional quantum Hall filling factors (e.g. $\nu_t=1/(2n+1)$ for the Laughlin states), a gapped topological phase can emerge in dLL with short range interaction, carrying a quantized Hall conductivity $\sigma_{xy}=\nu_t$.

In contrast when $\lambda>0$ and $N_e<N_{\phi}-N_{\delta}$, electrons occupy only the dLL, yielding $\nu_{dl}=N_e/\left(N_{\phi}-N_{\delta}\right)$, a gapped topological phase can also emerge when $\nu_{dl}=\nu_t$, contributing to the quantized Hall conductivity $\sigma_{xy}=\nu_t$. When $N_e\geq N_{\phi}-N_{\delta}$, the topological band becomes completely filled with $\nu_{dl}=1$, while the excess $N_e-(N_{\phi}-N_{\delta})$ electrons reside in the trivial bands. The total Hall conductivity will thus be non-quantized when the fermi surface does not lie within the band gap, from a sum of the quantized $\sigma_{xy}$ from the dLL and the non-quantized contribution from the dispersive bands\cite{supple}. The four different cases are summarized in Table.~\ref{Table1}.
\begin{table}[htbp]
    \centering
    \renewcommand{\arraystretch}{1.5}
    \begin{tabular}{ c | c | c | c | c | c }
    \hline\hline
         \textbf{Case} & $\lambda$  &  Condition & $\nu_{dl}$ & $\nu_l$ & $\sigma_{xy}$(${e^2}/{h}$) \\
         \hline\hline
          \textbf{I} & $\lambda<0$ & $N_e = N_{\delta}$ & 0 & $\nu_l=\frac{q}{p}$  & 0  \\
         \hline
         \textbf{II}  & $\lambda<0$ & $N_e>N_{\delta}$ & $\nu_t$ & $\nu_l=\frac{q+\nu_t\cdot(p-q)}{p}$ & $\nu_t$ \\
         \hline
         \textbf{III} & $\lambda>0$ & $N_e <N_{\phi}-N_{\delta}$ & $\nu_t$ & $\nu_l=\nu_t\cdot\frac{p-q}{p}$ & $\nu_t$ \\
         \hline
         \textbf{IV} & $\lambda>0$  & $N_e= N_{\phi}-N_{\delta}$ & 1 & $\nu_l=\frac{p-q}{p}$ & 1 \\
         \hline\hline
    \end{tabular}
    \caption{Four cases of conductivity and their corresponding filling factors on the LLL and decorated LL for small interaction as a perturbation.}
    \label{Table1}
\end{table}

An important feature here is that the Hall conductivity $\sigma_{xy}$ generally differs from $\nu_l$, the filling of the LL. A simple concrete example is to take $V_{int}$ as a short-range two-body interaction supporting a Laughlin state at $\nu_t=1/3$, and $N_{\delta}=N_{\phi}/2$ (i.e. $p=2,q=1$), with a single dispersive trivial band. In the regime when $\lambda<0$, the trivial band is partially filled while the dLL is empty for $\nu_l<1/2$, producing a gapless Fermi liquid phase with a non-quantized Hall conductivity. At $\nu_l=2/3$, the trivial band is fully filled while the dLL is one-third filled, realizing a gapped Laughlin phase with Hall conductivity $\sigma_{xy}=1/3$. In contrast, a positive lattice ($\lambda>0$) at $\nu_l=1/6$ yields the Laughlin state in the dLL with $\sigma_{xy}=1/3$, whereas the trivial band is empty. 

The phase diagram is schematic and may omit additional phases (e.g. the CDW or even superconducting phases in dispersive bands or dLL). There are two subtleties that demonstrates the richness of such systems. Firstly, for $q>1$, multiple dispersive bands with finite gaps can produce IQHE at fractional fillings in the weak-interaction limit, due to nontrivial Chern numbers of individual bands. Secondly, when the dLL is partially filled and dispersive bands fully filled with $\lambda<0$, inter-band interaction renormalize the dLL dispersion, potentially enhancing or suppressing topological phases\cite{supple}.

\textit{Interaction as the dominant energy scale--} Generally speaking, strong interaction tends to mix the dLL with the dispersive bands, as long as the interaction strength is comparable to the single particle band gaps. However, it is important to note that band mixing in LLs also strongly depends on the electron density, or the filling $\nu_l$ of the LL. It has been systematically studied before that for short-range interactions, mixing between different LLs can be completely suppressed when the electron filling is small, no matter how strong the interaction is~\cite{scalefree}. As will be shown, this is also the case for the dLLs, leading to unexpected topological phases gapped by both the interaction and the lattice potential.

We again start with the general cases, with short-range $V_{\text{int}}$ giving gapped topological phases for undecorated LLL at filling factor $\nu_t$ and $|\lambda|\ll 1$ in Eq.(\ref{vint}). In this case there will be strong mixing of the dLL and the trivial bands due to the interaction, and the delta potentials serve as a small perturbation. We naturally expect the topological phase $\nu_l=\nu_t$ to be robust. This is also true for other topological phases, such as the Jain hierarchical states, that can be supported by the short-range interaction, as long as the strength of the delta potentials is smaller than the incompressibility gap.

Interestingly, when $\lambda>0$ and $\nu_{dl}\le\nu_t$ and $V_{\text{int}}$ is the model Hamiltonian (i.e. the Trugmann-Kivelson interaction\cite{Trugmann-Kivelson}) for $\nu_t=1/(2n+1)$, all electrons are confined within the dLL. In particular an incompressible phase can form with $\nu_{dl}=\nu_t$ giving quantized $\sigma_{xy}=\nu_t$, with the gap supported by $\lambda$. It is worth highlighting that there will be no band mixing, no matter how strong the interaction is when $\nu_{dl}\leq \nu_t$. This is because all electrons can stay within dLL, forming many-body states with zero energy with respect to the model Hamiltonian. 
When $\nu_{dl}=\nu_t$ there is a unique zero energy state~(with center-of-mass degeneracy on the torus), which is gapped by both the lattice potential and the interaction. 

The case of $\lambda<0$ is more complicated. The dominance of the interaction implies strong mixing between the dLL and dispersive bands, likely forming a gapless phase\cite{intontrivialband1,intontrivialband2,intontrivialband3,intontrivialband4} (more detailed analysis will be carried out elsewhere). The cases discussed above are summarized in Table~\ref{Table2}.
\begin{table}[tbp]
    \centering
    \renewcommand{\arraystretch}{1.5}
    \begin{tabular}{ c | c | c | c | c | c }
    \hline\hline
         \textbf{Case} & $\lambda$ & Condition & $\nu_{dl}$ & $\nu_l$ & $\sigma_{xy}$(${e^2}/{h}$) \\
         \hline
         \textbf{i} & $\lambda \gtrless 0$ & $\nu_l=\nu_t$ & $\nu_t$ & $\nu_l=\nu_t$ & $\nu_t$ \\
         \hline
         \textbf{ii} & $\lambda>0$ & $N_e <N_{\phi}-N_{\delta}$ & $\nu_t$ & $\nu_l=\nu_t \cdot\frac{p-q}{p}$ & $\nu_t$ \\
         \hline
         \textbf{iii} & $\lambda<0$ & $N_e\leq \nu_t N_{\delta}$ & N. A. & $\nu_l\leq \nu_t\cdot \frac{q}{p}$ & non-quantized \\
         \hline\hline
    \end{tabular}
    \caption{Three cases of conductivity and their corresponding filling factors on the LLL and decorated LL when the interaction in Eq. (1) dominates.}
    \label{Table2}
\end{table}
\begin{table*}[t]
\centering
\renewcommand{\arraystretch}{1.4} % adjust row height for readability
\begin{tabular}{c|c|c|c|c|c||c|c|c|c|c|c}
\hline\hline
\multicolumn{6}{c||}{\textbf{moir\'e Chern bands}} &
\multicolumn{6}{c}{\textbf{Landau Levels}} \\
\hline
\textbf{Case} & $\nu_m$  & $\nu_C$ & $\nu_0$ & Phase &  $\sigma_{xy}$ (${e^2}/{h}$)  & $\nu_l$ & $\nu_{dl}$ & $\lambda$ & LL dual in Table.~\ref{Table1} and \ref{Table2} & Phase &  $\sigma_{xy}$ (${e^2}/{h}$) \\
\hline \hline
\textbf{1} & 1/2 & 0 & 1 & insulating phase & 0 & 1/2 & 0 & $\lambda<0$ & \textbf{I}, interaction  perturbs & insulating phase & 0 \\
\hline
\textbf{2} & 1/6 & 0 & 1/3 & CDW phase & 0 & 1/6 & 0 & $\lambda<0$ & \textbf{I}, interaction  perturbs & Fermi liquid & non-quantized\\
\hline
\textbf{3} & 2/3 & 1/3 & 1 & FCI phase & 1/3 & 2/3 & 1/3 & $\lambda<0$ & \textbf{IV}, interaction  perturbs & topological phase & 1/3 \\
\hline
\textbf{4} & 5/6 & 2/3 & 1 & FCI phase & 2/3 & 5/6 & 2/3 & $\lambda<0$ & N. A. &  -  & - \\
\hline
\textbf{5} & 1/3 & 1/3 & 1/3 & FCI phase & 1/3 & 1/3 & 1/3 & $\lambda \gtrless 0$ & \textbf{i},
interaction dominates & topological phase & 1/3 \\
\hline\hline
\end{tabular}
\caption{Comparison of examples between moir\'e Chern bands and LLs. $\nu_m$: total filling factor of two bands in moir\'e system. $\nu_C$:  filling factor of the Chern sub-band in moir\'e system. $\nu_0$: filling factor of the topologically trivial band in moir\'e system. $\nu_l$: filling factor of the entire LL. $\nu_{dl}$: filling factor of the decorated LL. 
}
\label{Table3}
\end{table*}

We again look at a specific example of short-range interaction $V_{\text{int}}$ supporting gapped $\nu_t=1/3$ Laughlin phase and $p=2$, $q=1$. In this case, two distinct regimes of Hall conductivity arise. At the filling factor $\nu_l=1/3$, the lattice potential induces strong band mixing between the dLL and dispersive bands for both positive and negative lattice potential, maintaining a quantized Hall conductivity $\sigma_{xy}=1/3$. A more subtle case appears when $\lambda > 0$ and $\nu_l = 1/6$, the lattice potential repels the electrons and they are entirely confined to the dLL, leading to the band mixing vanishing. This yields $\nu_{dl} = 1/3$, realizing a Laughlin phase with Hall conductivity $1/3$, in which the smallest gap is set by the lattice energy scale.

%{\color{red}Moreover, unlike in the lattice-dominated regime, all FQH states persist when interactions prevail for $\nu_{l}\geq \nu$. For example, at filling factor $\nu_l = 1 - \nu$, the ground states of the model Hamiltonians remain robust against the lattice perturbation: although the degeneracy is lifted through mixing between the trivial bands and the decorated LL, the underlying topological character of conjugate states survives for both $\lambda>0$ and $\lambda<0$.} For instance, considering $V_{{int}} = V_1$, $p=2$, $q=1$ and $\nu_l = 2/3$, although exact zero-energy ground states are absent, three quasi-degenerate ground states stabilized by the dominant interaction sustain a large FQH gap, yielding a quantized Hall conductivity of $\sigma_{xy} = \tfrac{2e^2}{3h}$, independent of the sign of $\lambda$. In addition, similar subspaces mixing arises in fractional quantum Hall phases that lack exact model Hamiltonians, such as the Jain hierarchical states, leading to a splitting of the ground states manifold while preserving the topological phase.
\begin{figure}[ht]
\centering
\includegraphics[width=1.0 \linewidth]{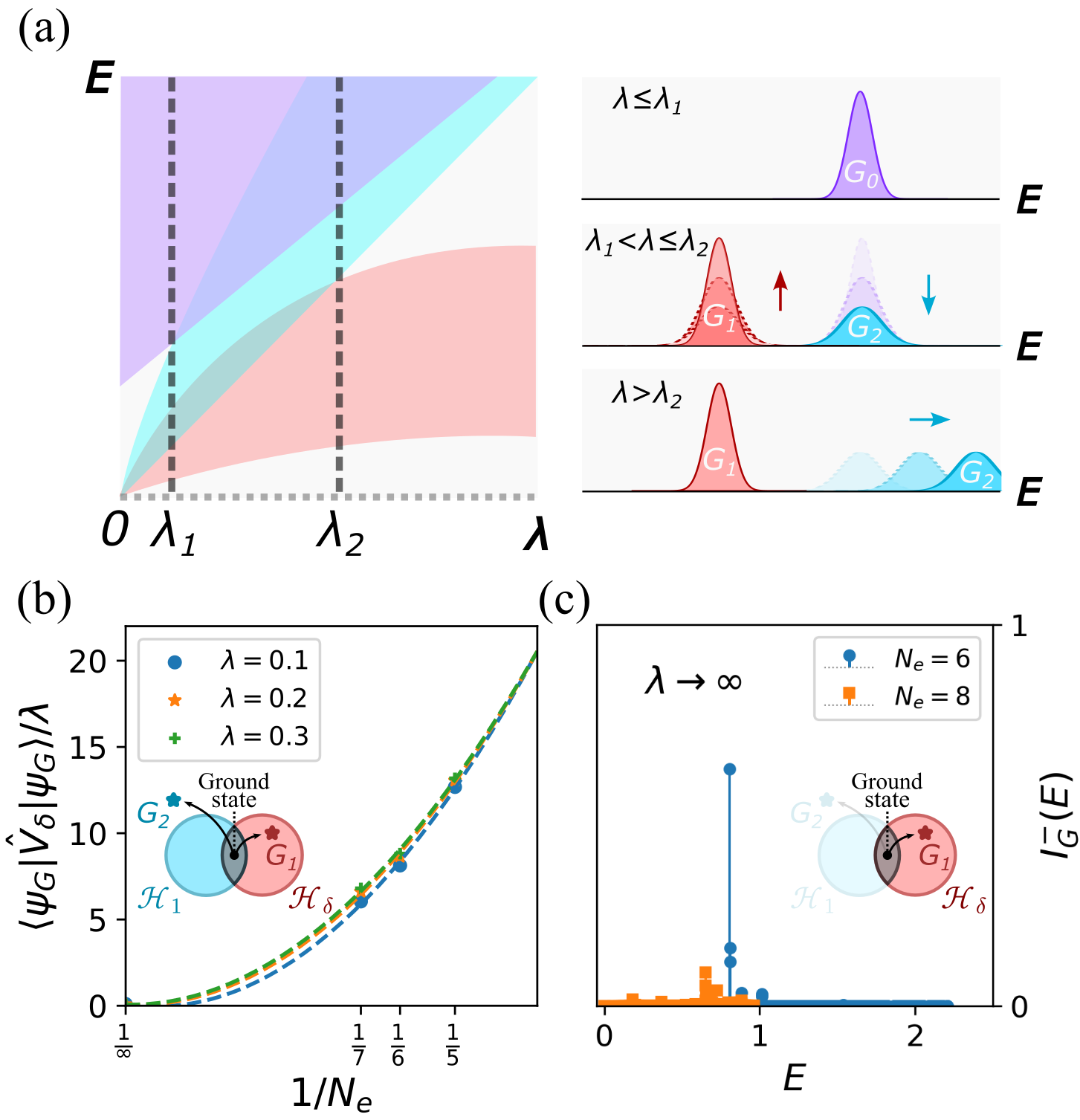}
\caption{\textbf{(a) Schematic of the spectrum evolution with $\lambda$ (left) and the graviton mode (GM) behavior (right).}
Using the $V_1$ pseudopotential at $\nu_l=1 / 6$ with $N_\delta=N_o / 2$ : for small $\lambda$, purple states lie outside the $V_1$ nullspace $\mathcal{H}_1$, red(blue) are Laughlin quasiholes inside(outside) the dLL $\mathcal{H}_\delta$, giving a single GM peak $G_0$. Increasing $\lambda$ mixes sectors and splits $G_0$ into two.
\textbf{(b) Finite size scaling of the GM energy for $\lambda > \lambda_2$.} The finite size trend indicates that the GM asymptotically resides within the dLL.
\textbf{(c) Spectral function of the dLL GM at large $\lambda$.} In the lattice-dominated regime, the GM peak rapidly decays with system size, indicating short-lived GMs in the dLL.
}
\label{fig4}
\end{figure}
\begin{figure}[h!]
\centering
\includegraphics[width=1.0 \linewidth]{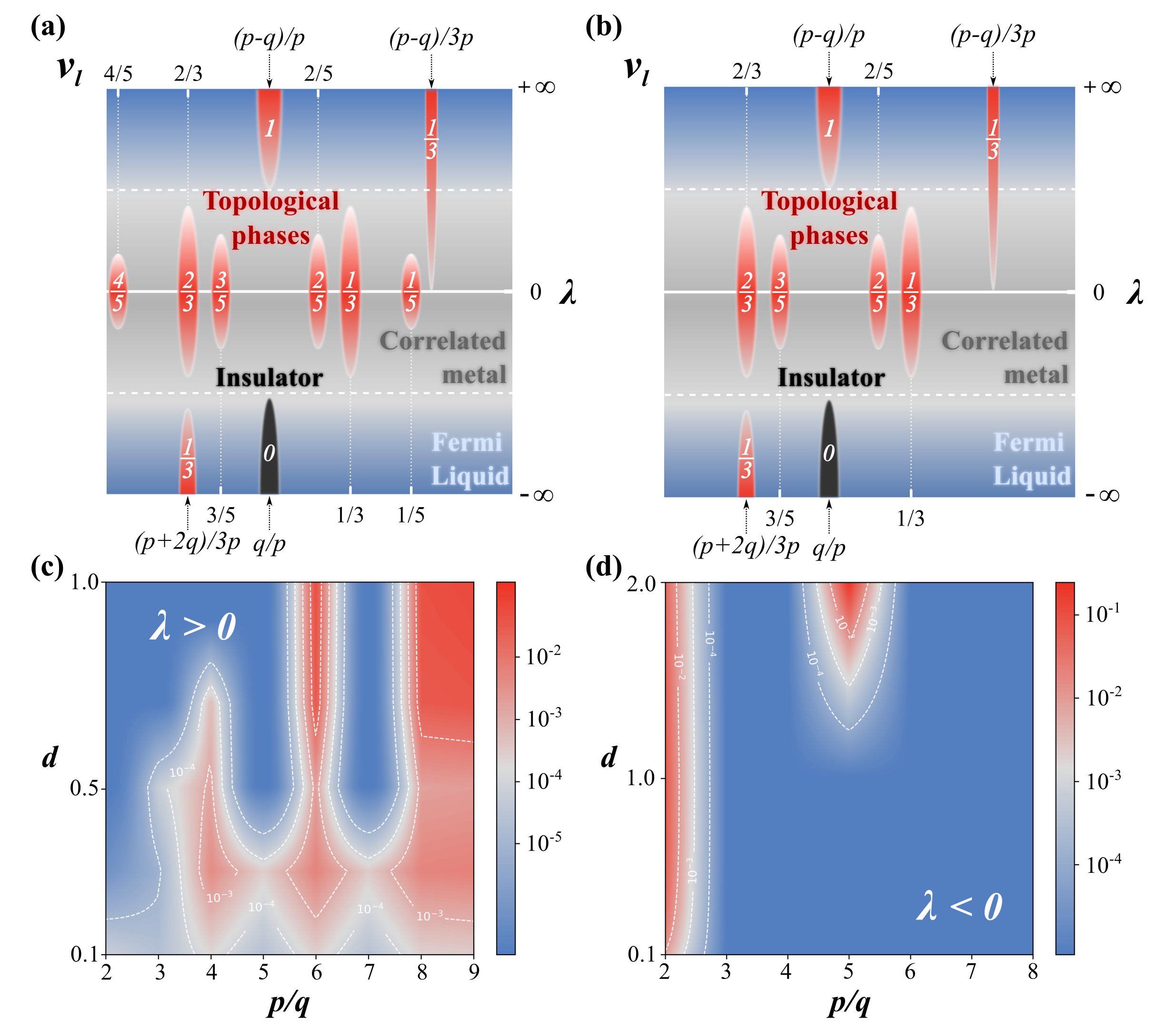}
\caption{\textbf{Phase diagram of the Hamiltonian in Eq. (\ref{vint})}. The blue, grey, red and black area indicate Fermi liquid, correlated metal, topological phases, and insulating phase, respectively. The values on the topological phases indicate the phase conductivity. (a) Phase diagram for Yukawa interaction,  $V(\bm{r})=\frac{e^{-|\bm{r}|/d}}{|\bm{r}|}$. (b) Phase diagram for interaction $V_1$. (c) Phase diagram for Yukawa interaction when $\lambda>0$, $\nu_{dl}=1/3$ and dispersive bands are empty. The scale of the color shows the scale of the FQH gap. (d) Same as (c) but for $\lambda<0$, so $\nu_{dl}=1/3$ and dispersive bands are fulled filled~\cite{footnote3}.
}
\label{fig5}
\end{figure}

\textit{Neutral excitations in decorated LLs--} We now turn to the study of gapped excitations in dLLs and examine how they differ from their counterparts in conventional LLs. Of particular interest are the geometric collective modes that arise from fluctuations of the metric associated with each conformal Hilbert space (CHS), commonly referred to as graviton modes (GMs) \cite{GGM1,GGM2,GGM3}. These modes carry spin $S = 2$ with well-defined chiralities. In conventional FQH phases, GMs appear as long-lived neutral excitations embedded in the excitation continuum, as observed in experiments \cite{GGM4}, and provide a dynamical probe of the emergent geometric degrees of freedom in the underlying many-body state. In the dLL, however, both the dynamics and the lifetime of the GM can be modified by the interplay between interaction effects and the quantum geometries tuned by single-particle potentials.

Since the FQH phases at $\nu_l = 1/3$ in the LLL are well understood, we study the dLL at the same effective filling $\nu_{dl}=1/3$ using the $V_1$ pseudopotential, obtained by introducing $N_\delta = N_o/2$ delta potentials at $\nu_l = 1/6$ in the LLL. The GM remains fully chiral and invariant under tuning $\lambda$, lying entirely outside the $V_1$ nullspace, while part of its spectral weight may project outside the dLL. As a result, the GM can in principle exhibit two spectral peaks: one associated with metric fluctuations of the $V_1$ CHS restricted to the dLL, and another associated with the CHS of the dLL. As illustrated in Fig.~\ref{fig4}(a), in the small–$\lambda$ limit, most of the GM weight resides in a single peak $G_0$. As $\lambda$ increases,  a new GM peak $G_1$ develops within the dLL and becomes the dominant geometric mode. In the large–$\lambda$ regime, $G_1$ stabilizes as the intrinsic dLL GM, whereas the $G_2$ peak outside dLL is pushed to higher energies.

Fig.~\ref{fig4}(b) shows the finite size scaling of the variational energy of the GM with respect to the delta potentials. The trend indicates that, however, the GM may reside entirely within the dLL in the thermodynamic limit. Meanwhile, the GM energy exhibits a discrete jump when the system is tuned from the small–$\lambda$ to the large–$\lambda$ regime, reflecting the fact that the interaction energy scale changes from the LLL to the dLL~\cite{footnote2}. In the lattice-dominated regime, the spectral function in Fig.~\ref{fig4}(c) reveals that the dLL GM peak decays rapidly with increasing system size, indicating that the mode acquires a short lifetime due to scattering into nearby excitations. This behavior contrasts with the long-lived GM in LLs, and is instead consistent with GM dynamics in moir\'e Chern bands \cite{lifetime2}, where the density of states around the GM energy controls its decay. These results suggest that the GM in the dLL can only be observed experimentally if it is tuned out of the continuum to lower energies. A systematic investigation of possible tuning mechanisms and the role of electric and magnetic disorder potentials is left for future work.

\textit{Discussions and outlook--} As we mentioned in the beginning, the decorated LL, spanned by the exact zero-energy states of delta potentials, does not require delta potentials to be spatially arranged in a lattice. The universal properties of all gapped topological phases are robust even if the positions of the delta potentials are disordered, as long as there is a clear domination of the interaction or the single particle energy scale. This is a clear demonstration that these topological phases do not require rotational or (discrete) translational symmetry. 

Between the well-defined incompressible phases, our phase diagram in fig.~\ref{fig5} reveals broad intermediate regimes conjectured to be mostly characterized by gapless behavior.  These compressible regions are highly sensitive to microscopic details of the experimental realization, including the spatial distribution, strength and profile of the local potentials.  Unlike topological phases, these states lack universal quantization and may host rich competing orders.  Depending on parameters, the intermediate regimes could accommodate CDW ordering, composite Fermi liquid behavior, or percolation-driven metallic phases akin to those proposed in disordered Chern bands\cite{corelatedmetal1,corelatedmetal2,corelatedmetal3}.  Identifying the precise nature of these compressible states requires intensive numerical simulations and detailed experimental studies.

The dLL studied in this work can also reveal the underlying mechanisms of rather surprising fractional quantum anomalous Hall effect in certain moir\'e superlattice systems, as recently reported in ref.{\cite{sunkai}}. There the physics of two degenerate moire bands can be captured by the dLL with $p=2,q=1$, as summarized in Table.~\ref{Table3}. There is no external periodic electrostatic potential within the moire flat band, but an emergent periodic one-body potential is expected due to electron-electron interaction and the non-uniform Berry curvature distribution. Thus \emph{unlike} the dLL systems, the energy scale of the interaction and the emergent periodic potential are not independently tunable, though the strength of the one-body potential increases with electron density at the Hartree-Fock level. Consequently only a small number of gapped phases easily realizable in dLL can be realized in the carefully tuned moire systems, such as Case 1,3 and 5 in Table.~\ref{Table3}. It is interesting to note that for Case 4, a gapped topological phase is found in the moire systems, though it seems difficult to realize such a gapped phase with a square lattice of delta potentials in the dLL systems. We however expect such gapped phase to be supported in dLL with more careful tuning of interaction and one-body periodic potentials.

From an experimental standpoint, the robustness of dLLs open up a promising route to realize highly tunable topological platforms for exotic correlated phases in the two-dimensions. Decorated LLs can be engineered using electrostatic tips, substrate patterning or Coulomb blockade microscopy\cite{tip1,tip2,localexp1,localexp2,dllexp1,dllexp2,dllexp3}. The tuning parameters include the strength of lattice potential $\lambda$, the type of the lattice (e.g. with different discrete symmetry or $p,q$), or even disordered localized potentials. This is especially true for cases when $p$ is large (experimentally more feasible), splitting a single LL into $p$ bands with their physical properties dependent on the details of the one-body potential.  We expect rich phases of the 2D quantum fluids or solids to emerge (e.g. correlated mental, CDW phase, and even superconductivity) that deserve systematic future studies.

\textit{Acknowledgement--} We would like to acknowledge useful discussions with Kai Sun, Qianhui Xu, Ha Quang Trung, and Chenxin Jiang. This work is supported by the National Research Foundation, Singapore under the NRF Fellowship Award (NRF-NRFF12-2020-005), Singapore Ministry of Education (MOE) Academic Research Fund Tier 3 Grant (No. MOE-MOET32023-0003) “Quantum Geometric Advantage”, and Singapore Ministry of Education (MOE) Academic Research Fund Tier 2 Grant (No. MOE-T2EP50124-0017).

\onecolumngrid
\pagebreak
\widetext
\begin{center}
\textbf{\large Supplementary materials for ``Emergence and transition of incompressible phases in decorated Landau levels''}

\end{center}
%%%%%%%%%% Merge with supplemental materials %%%%%%%%%%
%%%%%%%%%% Prefix a "S" to all equations, figures, tables and reset the counter %%%%%%%%%%
\setcounter{equation}{0}
\setcounter{figure}{0}
\setcounter{table}{0}
\setcounter{page}{1}
\makeatletter
\maketitle
\renewcommand{\theequation}{S\arabic{equation}}
\renewcommand{\thefigure}{S\arabic{figure}}
\renewcommand{\thetable}{S\arabic{table}}
\renewcommand{\bibnumfmt}[1]{[S#1]}
\renewcommand{\citenumfont}[1]{S#1}

\section{ magnetic translation operators and splitting of a single LL into decorated LL and trivial bands}

Here we give two definitions of magnetic translation operators and their relationship. Then we find the  magnetic translation operator which commutes with the lattice potential, thus show how a single LL can be divided into two groups, the decorated LL and trivial bands.

One definition of the magnetic translation operator describes the displacement in real space, 
\begin{equation}
    \hat{\widetilde{t}}(\bm{R})=exp(-\frac{i}{\hbar}\bm{R}\cdot \hat{\bm{\Pi}}),
\end{equation}
where 
\begin{equation}
    \hat{\bm{\Pi}}=\hat{\bm{p}}+e\bm{A}(\bm{r}),
\end{equation}
in which $\bm{A}$ is the vector potential.

And the other definition of magnetic translation operator describes the displacement in momentum space,
\begin{equation}
    \hat{T}(\bm{q})=exp(i\bm{q}\cdot \hat{\bar{\bm{R}}}),
\end{equation}
where $\hat{\bar{\bm{R}}}$ is the guiding center coordinate.

Let's decompose the particle coordinate $\hat{\bm{r}}$ into guiding center and cyclotron parts,
\begin{equation}
    \hat{\bm{r}}=\hat{\bar{\bm{R}}}+\hat{\bm\eta}.
\end{equation}
Thus $\hat{\bm{\Pi}}$ can be written as:
\begin{equation}
    \hat{\bm{\Pi}}=\frac{\hbar}{\ell_B^2}\hat{z}\times \hat{\bm\eta}.
\end{equation}
Thus $\hat{\widetilde{t}}(\bm{R})$ can be written as:
\begin{equation}
\begin{split}
    \hat{\widetilde{t}}(\bm{R}) & = exp\left (-\frac{i}{\hbar}\bm{R}\cdot  \hat{\bm{\Pi}}\right)\\
    & =exp\left( -\frac{i}{\hbar}\bm{R}\cdot \frac{\hbar}{\ell_B^2}(\hat{z}\times \hat{\bm{\eta}})\right)\\
    & = exp\left(-i\frac{\bm{R}\cdot (\hat{z}\times \hat{\bm{\eta}})}{\ell_B^2}\right).
\end{split}
\end{equation}
With the identity
\begin{equation}
    \bm{R}\cdot (\hat{z}\times \hat{\bm{\eta}})=-\hat{\bm{\eta}}\cdot (\hat{z}\times \bm{R}).
\end{equation}
We have:
\begin{equation}
    \hat{\widetilde{t}}(\bm{R})=exp\left( i\frac{\hat{\bm{\eta}}\cdot (\hat{z}\times \bm{R})}{\ell_B^2}\right).
\end{equation}
Without loss of generality, we set,
\begin{equation}
    \bm{q}(\bm{R})=\frac{\hat{z}\times \bm{R}}{\ell_B^2}.
\end{equation}
Hence,
\begin{equation}
    \hat{\widetilde{t}}(\bm{R})=exp(i\bm{q}(\bm{R})\cdot \hat{\bm{\eta}})
\end{equation}
Since,
\begin{equation}
    exp(i\bm{q}\cdot \hat{\bm{r}})=exp(i\bm{q}\cdot \hat{\bar{\bm{R}}})exp(i\bm{q}\cdot \hat{\bm{\eta}}),
\end{equation}
\begin{equation}
    exp(i\bm{q}\cdot \hat{\bm{r}})=exp(i\bm{q}\cdot \hat{\bar{\bm{R}}})\hat{\widetilde{t}}(\bm{r}),
\end{equation}
Thus,
\begin{equation}
    \hat{\widetilde{t}}(\bm{r})=exp(i\bm{q}\cdot \hat{\bar{\bm{R}}})exp(-i\bm{q}\cdot \hat{\bm{\eta}}).
\end{equation}
So we here claim the relationship between the two magnetic translation operators is: the projection of $\hat{\widetilde{t}}(\bm{r})$ onto a single Landau Level $\hat{t}(\bm{R})$ is the product of form factor and $\hat{T}(\bm{q})$. That is,
\begin{equation}
    \begin{split}
        \hat{t}(\bm{R}) & =P_{NLL}\hat{\widetilde{t}}(\bm{r})P_{NLL}\\
         & = \sum_N \ket{N}\bra{N} exp(i\bm{q}\cdot \hat{\bar{\bm{R}}})exp(-i\bm{q}\cdot \hat{\bm{\eta}}) \ket{N}\bra{N} \\
         & = \sum_N F_{NLL}\ket{N} exp(i\bm{q}\cdot \hat{\bar{\bm{R}}}) \bra{N}\\
         & = F_{NLL}exp(i\bm{q}\cdot \hat{\bar{\bm{R}}})\\
         & = F_{NLL} \hat{T}(\bm{q}).
    \end{split}
\end{equation}

The commutation between the kinetic Hamiltonian and a series of magnetic translation operators is as follows,
\begin{equation}
    \left[H_{\bm{k}}, T_{\bm{R}} \right]=0,
\end{equation}
where $\bm{R}=m\bm{a}_1'+n\bm{a}_2'$. Here $\bm{a}_1'$ and $\bm{a}_2'$ can be taken in any way as soon as they span a magnetic unit flux by $|\bm{a}_1' \times \bm{a}_2'|=2\pi$; m and n are integers. The commutation allows basis of a single LL to be labeled by the Bloch wave, obeying:
\begin{equation}
    \hat{t}(\bm{a}_1')\ket{k_x,k_y}=e^{\frac{i2\pi k_x}{N_x}}\ket{k_x,k_y},
\end{equation}
\begin{equation}
    \hat{t}(\bm{a}_2')\ket{k_x,k_y}=e^{\frac{i2\pi k_y}{N_y}}\ket{k_x,k_y}.
\end{equation}

Let's consider a delta potential lattice, whose lattice vectors are $\bm{a}_1$ and $\bm{a}_2$, 
\begin{equation}
    |\bm{a}_1 \times \bm{a}_2|=\frac{p}{q}2\pi.
\end{equation}
Since
\begin{equation}
    \hat{t}(\bm{R}_1)\hat{t}(\bm{R}_2)=e^{-i\frac{eB}{\hbar}(\bm{R}_1 \times \bm{R}_2)\cdot \hat{z}}\hat{t}(\bm{R}_2)\hat{t}(\bm{R}_1),
\end{equation}
we now define a series of magnetic translation operators $\hat{t}_{\bm{R}_{mn}}$ where $\bm{R}_{mn}=m \bm{a}_1+q\cdot n \bm{a}_2$ without loss of generality such that 
\begin{equation}
    \left[\hat{V}_{local}, \hat{t}_{\bm{R}_{mn}} \right]=0,
\end{equation}
where $\hat{V}_{local}$ is the lattice potential in the main text. Now we define two magnetic translation operators $\hat{t}_1$ as $\hat{t}_{\bm{R}}$ when $\bm{R}=\bm{a}_1$ and $\hat{t}_2$ as $\hat{t}_{\bm{R}}$ when $\bm{R}=q\bm{a}_2$, and take $\bm{a}_1=\bm{a}_1'$, $\bm{a}_2=\frac{p}{q}\bm{a}_2'$. Thus,

\begin{equation}
\begin{split}
    & \hat{t}_1=\hat{t}(\bm{a}_1'),\\
    & \hat{t}_2=\hat{t}(q\bm{a}_2) = \hat{t}(p\bm{a}_2')=\left( \hat{t}(\bm{a}_2') \right)^p .
\end{split}
\end{equation}
So $\bm{R}_{mn}=m \bm{a}_1'+p\cdot n\bm{a}_2'$.  Therefore,
\begin{equation}
\begin{split}
    & \hat{t}_1 \ket{k_x,k_y}= e^{\frac{i2\pi k_x}{N_x}}\ket{k_x,k_y},\\
    & \hat{t}_2 \ket{k_x,k_y}=e^{\frac{i2\pi k_y \cdot p}{N_y}}\ket{k_x,k_y}.
\end{split}
\end{equation}
Thus a series of states $\ket{k_x, k_y=k_{y,0}+t\frac{N_y}{p}}$ share the same crystal momentum $\frac{2\pi \cdot k_{y,0}}{N_y/p}$, where $k_{y,0}=0,1,\cdots \frac{N_y}{p}-1$, by
\begin{equation}
\begin{split}
    \hat{t}_2 \ket{k_x,k_y} & =exp\left( \frac{i2\pi \cdot p (k_{y,0}+t\cdot \frac{N_y}{p})}{N_y}  \right)\ket{k_x,k_y}\\
    & =exp\left(\frac{i2\pi\cdot k_{y,0}}{N_y/p} \right)\ket{k_x,k_y},
\end{split}
\end{equation}
where $t=0,1,\cdots p-1$. 

Therefore, the eigenstates of lattice potential can be obtained by diagonalising the matrix calculated with the basis in the same crystal momentum sector, which share the same crystal momentum. In the momentum space, the lattice potential in the main text can be written as
\begin{equation}
        \hat{V}_{local} =\sum_{i} \sum_{\bm{q} }e^{i\bm{q}(\hat{\bm{r}}-\bm{r}_i)} = \sum_{\bm{q}} e^{i\bm{q}\hat{\bm{r}}}\sum_{i}e^{i\bm{q}\bm{r}_i}.
\end{equation}
Tthe projection of local potential lattice Hamiltonian onto LLL is:
\begin{equation}
    \hat{V}_{local}=\sum_{\bm{q}} e^{i\bm{q}\hat{\bar{\bm{R}}}}F_{NLL}(\bm{q})\sum_{i}e^{i\bm{q}\bm{r}_i}.
\end{equation}
Writing this lattice potential Hamiltonian into the second quantization form:
\begin{equation}
\hat{V}_{local}=\sum_{\bm{k},\bm{k}'}V_{\bm{k},\bm{k}'}c^{\dagger}_{\bm{k}'}c_{\bm{k}},
\end{equation}
the matrix element is given by:
\begin{equation}
\begin{split}
\bra{\bm{k}}\hat{V}_{local}\ket{\bm{k}'} & =\bra{\bm{k}}\sum_{\bm{q}} e^{i\bm{q}\cdot \hat{\bar{\bm{R}}}}F_{NLL}(\bm{q})\sum_{i}e^{i\bm{q}\cdot \bm{r}_i}\ket{\bm{k}'}\\
& =\sum_{\bm{q}} F_{NLL}(\bm{q})\sum_{i}e^{i\bm{q}\cdot \bm{r}_i}\bra{\bm{k}}e^{i\bm{q}\cdot \hat{\bar{\bm{R}}}}\ket{\bm{k}'} .\\
\end{split}
\end{equation}
Notice the periodic boundary condition gives:
\begin{equation}
    \ket{\bm{q}+\bm{b}}=\eta_{\bm{b}} e^{-\frac{i}{2}\bm{b}\times \bm{k}}\ket{\bm{k}},
\end{equation}
in which $\eta_{\bm{b}}=1$ if $\bm{b}/2$ is reciprocal lattice and  $\eta_{\bm{b}}=-1$ if $\bm{b}/2$ is not. And,
\begin{equation}
    e^{i\bm{    }\hat{\bar{\bm{R}}}}\ket{\bm{k}}=e^{\frac{i}{2}\bm{ q}\times \bm{k}}\ket{\bm{k}+\bm{q}}.
\end{equation}
\begin{equation}
    \begin{split}
        & \bra{\bm{k}}e^{i\bm{q}\cdot \hat{\bar{\bm{R}}}}\ket{\bm{k}'} \\ 
        &=\bra{\bm{k}}e^{\frac{i}{2}\bm{    }\times \bm{k}'}\ket{\bm{k}'+\bm{   q}}\\
        & \propto \delta'(\bm{k},\bm{k}'+\bm{q   }), \\
    \end{split}
\end{equation}
in which $\delta'(\bm{k},\bm{k}'+\bm{q})=1$ if $\bm{k}-\bm{k}'-\bm{q}=c\bm{b}_1+d\bm{b}_2$ where c and d are integers.
Saying $\bm{k}=\bm{k}'+\bm{q}+\bm{b}$,
\begin{equation}
    \begin{split}
        \bra{\bm{k}}e^{\frac{i}{2}\bm{  }\times \bm{k}'}\ket{\bm{k}'+\bm{q}}& = e^{\frac{i}{2}\bm{q}\times \bm{k}'}\bra{\bm{k}} \eta_{\bm{b}} e^{\frac{i}{2}\bm{b}\times (\bm{k}'+\bm{q})}\ket{\bm{k}'+\bm{q}+\bm{b}}\\
        & = e^{\frac{i}{2}\bm{q}\times \bm{k}'} \eta_{\bm{b}} e^{\frac{i}{2}\bm{b}\times (\bm{k}'+\bm{q})} \delta(\bm{q},\bm{k}-\bm{k}'-\bm{b})\\
        & = \eta_{\bm{b}} e^{\frac{i}{2}(\bm{k}-\bm{k}'-\bm{b})\times \bm{k}'}e^{\frac{i}{2}\bm{b}\times (\bm{k}-\bm{b})}\\
        & = \eta_{\bm{b}}e^{\frac{i}{2}\bm{k}\times \bm{k}'}e^{\frac{i}{2}(\bm{k}+\bm{k}')\times \bm{b}}.
    \end{split}
\end{equation}
Thus,
\begin{equation}
    \bra{\bm{k}}\hat{V}_{local}\ket{\bm{k}'}=\sum_{\bm{b}} \eta_{\bm{b}} e^{\frac{i}{2}\bm{k}\times \bm{k}'}e^{\frac{i}{2}(\bm{k}+\bm{k}')\times \bm{b}} \sum_{i}e^{i(\bm{k}-\bm{k}'-\bm{b})\cdot \bm{r}_i} F_{NLL}(\bm{k}-\bm{k}'-\bm{b}).
\end{equation}

\section{ Berry curvature of the decorated LL and dispersive Chern bands}

Here we show calculations of Berry curvature on the decorated LL.

The Berry connection can be defined as
\begin{equation}
    A_{\bm{k}}^a=-i\braket{u_{\bm{k}}|\partial_{\bm{k}}^au_{\bm{k}}}, 
\end{equation}
where $\ket{u_{\bm{k}}}$ is the Bloch part of the wavefunction,
\begin{equation}
    \psi_{\bm{k}}=u_{\bm{k}}e^{i\bm{k}\cdot \bm{r}}.
\end{equation}
Considering the grid in crystal momentum space, the Berry phase of one piece of the grid can be calculated by the Wilson loop,
\begin{equation}
    \begin{split}
    e^{i\gamma} & = e^{\oint_C d\bm{k}\braket{u_{\bm{k}}|\nabla|u_{\bm{k}}}}\\
    & = e^{\sum_{t=1}^4} \delta\bm{k} \braket{u_{\bm{k}}^t|\nabla|u_{\bm{k}}^{t}}\\
     & = \prod_{t=1}^4 \left(1+ \delta\bm{k} \braket{u_{\bm{k}}^t|\nabla|u_{\bm{k}}^{t}}\right) \\
      & = \prod_{t=1}^4 \left(\braket{u_{\bm{k}}^t|u_{\bm{k}}^{t}}+ \delta\bm{k} \braket{u_{\bm{k}}^t|\nabla|u_{\bm{k}}^{t}}\right)\\
       & = \prod_{t=1}^4 \left(\braket{u_{\bm{k}}^t|u_{\bm{k}}^{t+1}}\right),\\
    \end{split}
\end{equation}
where $C$ is the integration path of the small piece of grid in momentum space, and  $u_{\bm{k}}^t$ denote the passing points. Four passing points are taken as $u_{\bm{k}}$, $u_{\bm{k}+\delta k\cdot \bm{b}_1}$, $u_{\bm{k}+\delta k\cdot \bm{b}_1+\delta k \cdot \bm{b}_2}$, and $u_{\bm{k}+\delta k\cdot \bm{b}_2}$, respectively, where $\bm{b}_1$ and $\bm{b}_2$ are reciprocal lattice vectors. Since the Berry phase is also the surface integration of Berry curvature, thus it can be approximately seen as the Berry curvature around the point that the small piece of grid stands, that is,
\begin{equation}
    \mathcal{B}(\bm{k}) \approx \gamma(\bm{k}) = Arg\left(\prod_{t=1}^4 \left(\braket{u_{\bm{k}}^t|u_{\bm{k}}^{t+1}}\right) \right).
\end{equation}
The overlap of Bloch parts of LL orbitals can be given by,
\begin{equation}
    \braket{u_{\bm{k}}|u_{\bm{k}+\bm{q}}}=\bra{\bm{k}}e^{i\bm{q}\cdot \hat{\bm{r}}}\ket{\bm{k}+\bm{q}}=f_{\bm{b}}^{\bm{k},\bm{k}+\bm{q}},
\end{equation}
where
\begin{equation}
    f_{\bm{b}}^{\bm{k},\bm{k}'}=F_{NLL}(\bm{k}-\bm{k}'-\bm{b})\eta_{\bm{b}}e^{\frac{i}{2}\bm{k}\times \bm{k}'}e^{\frac{i}{2}(\bm{k}+\bm{k}')\times \bm{b}}.
\end{equation}

With the lattice potentials, the Berry connection can be modified into,
\begin{equation}
    A_{\bm{k}}^a=-i\braket{v_{\bm{k}}|\partial_{\bm{k}}^av_{\bm{k}}}.
\end{equation}
According to previous supplementary materials, the eigenstates of the lattice potential are still Bloch states, denoted $\phi_{\bm{k}}(\bm{r})$, obeying
\begin{equation}
    \phi_{\bm{k}}(\bm{r})=v_{\bm{k}}(\bm{r})e^{i\bm{k}\cdot \bm{r}},
\end{equation}
in which $v_{\bm{k}}(\bm{r})$ is the Bloch part of the wavefunction. With the same definition of Berry Curvature and Berry phase, we only need to solve the overlap between the Bloch parts of eigenstates of the lattice potential,
\begin{equation}
    \braket{v_{\bm{k}}|v_{\bm{k}+\bm{Q}}}=\bra{\phi_{\bm{k}}}e^{-i \bm{k}\cdot \hat{\bm{r}}} e^{i (\bm{k}+\bm{Q})\cdot \hat{\bm{r}}}\ket{\phi_{\bm{k}+\bm{Q}}}=\bra{\phi_{\bm{k}}}e^{i \bm{Q}\cdot \hat{\bm{r}}} \ket{\phi_{\bm{k}+\bm{Q}}},
\end{equation}
in which $\bm{Q}$ can be taken as $\pm \bm{b}_1/N_x$ or $\pm \bm{b}_2/N_y$. Taking the the example $p=2$ and $q=1$ without loss of generality, and $\bm{b}_1$ and $\bm{b}_2$ as the reciprocal lattice vectors of the lattice potential, 
$\bm{b}_1'$ and $\bm{b}_2'$ can be constructed as the reciprocal lattice vectors of the magnetic unit, by $\bm{b}_1'=\bm{b}_1$, and $\bm{b}_2'=2\bm{b}_2$. The eigenstates of lattice potential can be written into linear combination of basis which share the same crystal momentum,
\begin{equation}
    \phi_{\bm{k}}(\bm{r})=c_{\bm{k}}\psi_{\bm{k}}(\bm{r})+c_{\bm{k}+\bm{b}_2}\psi_{\bm{k}+\bm{b}_2}(\bm{r}),
\end{equation}
in which the coefficients are numerically computed by diagonalising the matrix within the same momentum sector, as shown in eq. S32.
Therefore,
\begin{equation}
    \braket{v_{\bm{k}}|v_{\bm{k}+\bm{Q}}}=c_{\bm{k}}^{*}c_{\bm{k}+\bm{Q}}f_{\bm{b}}^{\bm{k},\bm{k+Q}}+c_{\bm{k}+\bm{b}_2}^{*}c_{\bm{k}+\bm{Q}+\bm{b}_2}f_{\bm{b}}^{\bm{k}+\bm{b}_2,\bm{k+Q}+\bm{b}_2}.
\end{equation}
For a more general case, the eigenstates of the lattice potential can be written into this form,
\begin{equation}
    \phi_{\bm{k}}(\bm{r})=\sum_{s=0}^{N_{\phi}/N_{\delta}-1} c_{\bm{k}+s\cdot \bm{b}_2}\psi_{\bm{k}+s\cdot \bm{b}_2}(\bm{r}).
\end{equation}
And,
\begin{equation}
    \braket{v_{\bm{k}}|v_{\bm{k}+\bm{Q}}}=\sum_{s=0}^{N_{\phi}/N_{\delta}-1}c_{\bm{k}+s\cdot\bm{b}_2}^{*}c_{\bm{k}+\bm{Q}+s\cdot \bm{b}_2}f_{\bm{b}}^{\bm{k}+s\cdot \bm{b}_2,\bm{k+Q}+s\cdot \bm{b}_2},
\end{equation}
in which $\bm{b}_1'=\bm{b}_1$ and $\bm{b}_2'=s\cdot \bm{b}_2$ without loss of generality.

Hence,
\begin{equation}
    \mathcal{B}(\bm{k}) \approx \gamma(\bm{k}) = Arg\left(\prod_{t=1}^4 \left(\braket{v_{\bm{k}}|v_{\bm{k}+\bm{Q}^t}}\right) \right).
\end{equation}
Four passing points are taken as $v_{\bm{k}}$, $v_{\bm{k}+1/N_x\cdot \bm{b}_1}$, $v_{\bm{k}+1/N_x\cdot \bm{b}_1+1/N_y\cdot \bm{b}_2}$, and $v_{\bm{k}+1/N_y\cdot \bm{b}_2}$, respectively. And four $\bm{Q}^t$ are taken as $1/N_x\cdot \bm{b}_1$, $1/N_y\cdot \bm{b}_2$, $-1/N_x\cdot \bm{b}_1$ and $-1/N_y\cdot \bm{b}_2$, respectively.

\begin{figure}[ht]
\centering
\includegraphics[width=0.65 \linewidth]{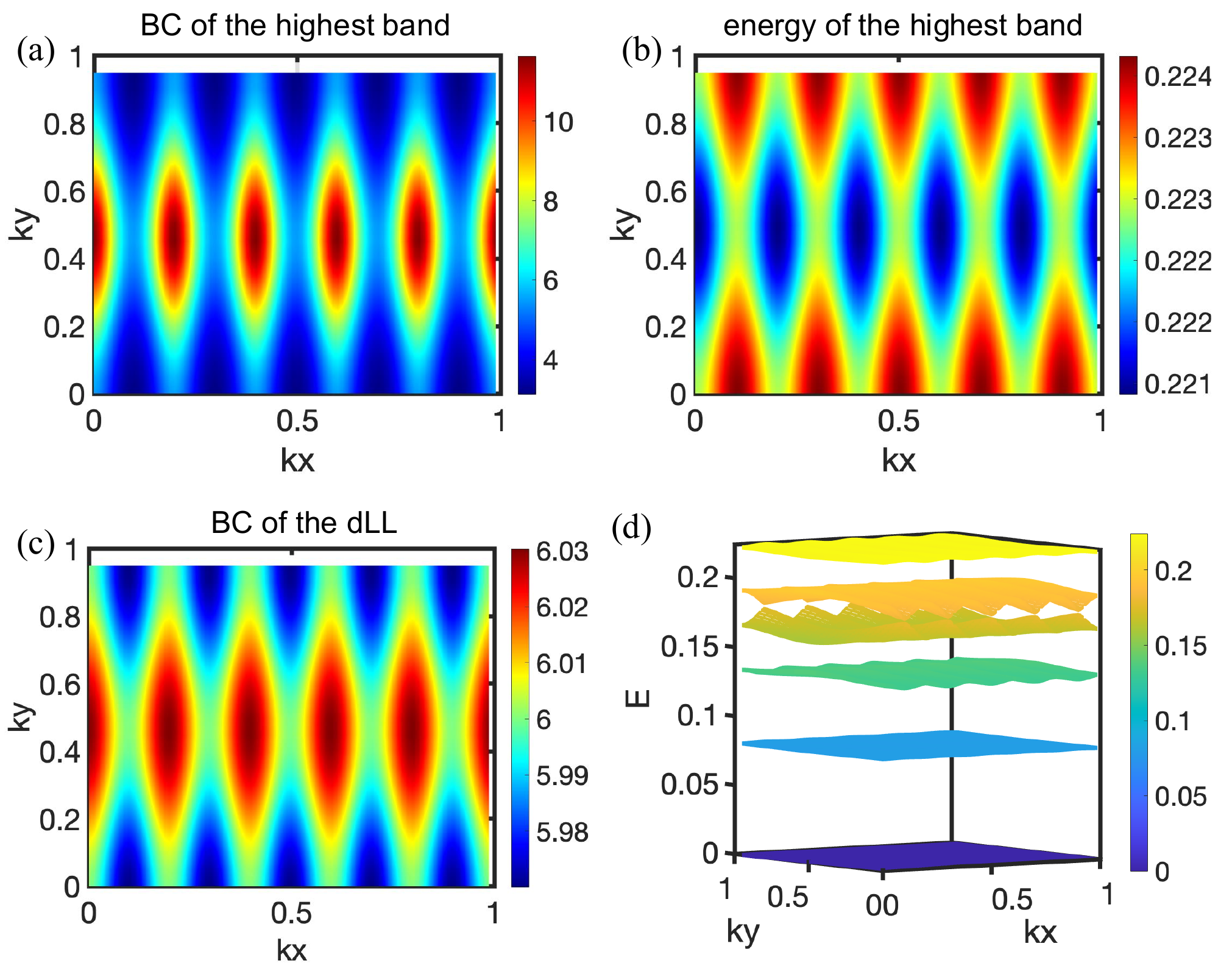}
\caption{The energy dispersion and Berry Curvature distribution of $p=6$ and $q=5$ case. BC: Berry Curvature. (a) Berry Curvature of the highest dispersive band. (b) Energy dispersion of the highest band. (c) Berry Curvature of the lowest band (dLL). (d) Spectra of $p=6$ and $q=5$, in which the highest dispersive band has Chern number 1.}
\label{figs4}
\end{figure}

Even though it is similar to the Landau level in many aspects, decorated LL has a non-uniform Berry Curvature as shown in fig.~\ref{figs4} due to the insertion of lattice potential, which turns on rich physics. For example, on the dLL, pure Coulomb interaction is not sufficient to give a Laughlin state when $p=2$ and $q=1$, as shown later. Moreover, since there are $q$ dispersive bands, even though they form a trivial band as a whole, some of dispersive bands can host non-trivial Chern number, with a gap away other bands. For example, we consider the case $p=6$, $q=5$, and the lattice is arranged as a square lattice without loss of generality. Thus we have 6 bands totally, the lowest exact zero-energy band is the dLL, and the other 5 bands are dispersive as shown in fig.~\ref{figs4}(d). Isolated from other bands, the highest band has a slight energy dispersion and heavily non-uniform Berry Curvature as shown in fig.~\ref{figs4}(b) and (a), and host a Chern number 1. Interestingly, the energy dispersion and Berry curvature distribution of both dLL and dispersive bands have similar profiles in real space, indicating that they inherit their geometric structure from the same Landau level subject to periodic electrostatic modulation, in which fig.~\ref{figs4}(c) is the Berry Curevature distribution of the dLL.

\section{ dLL splitting induced by realistic one-body potential}

\begin{figure}[ht]
\centering
\includegraphics[width=0.65 \linewidth]{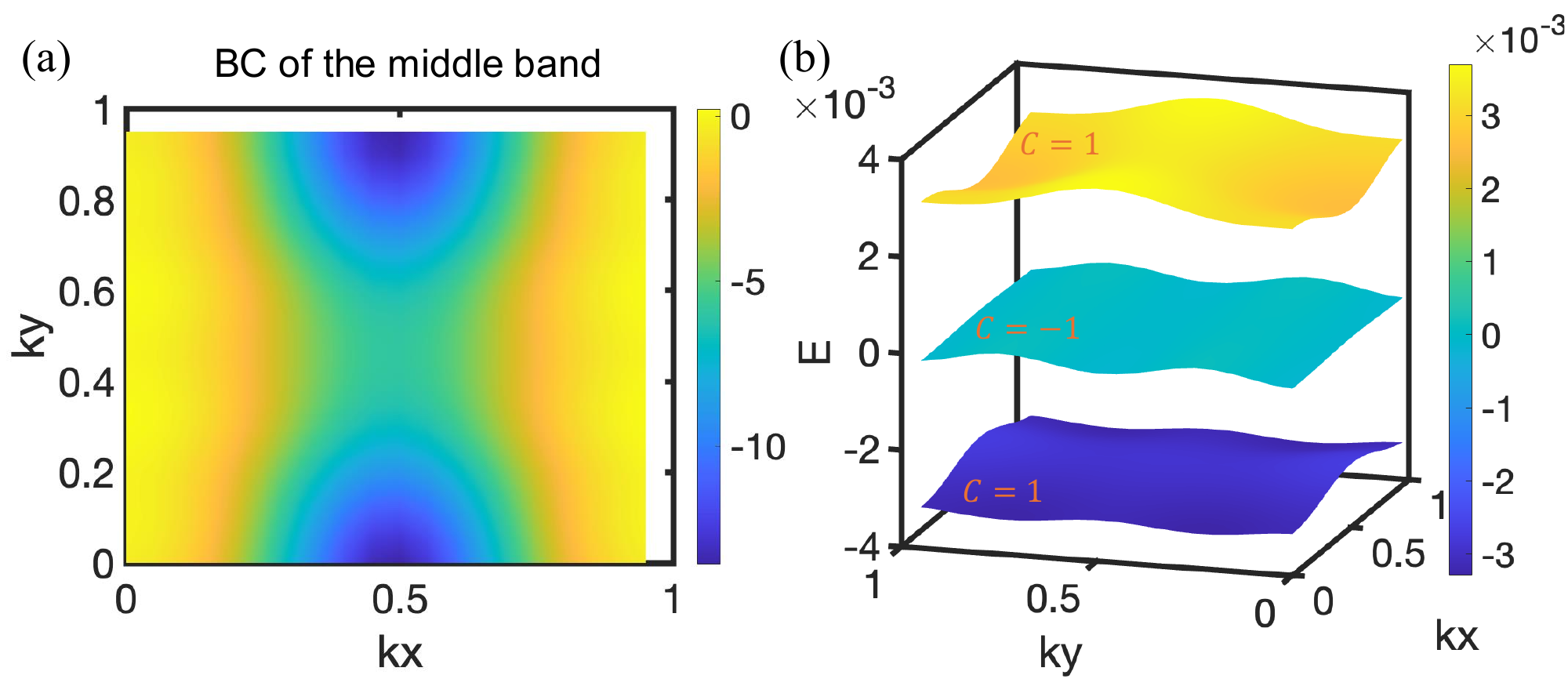}
\caption{(a) The Berry Curvature distribution of the middle band corresponding to $p=4$ and $q=1$ case after truncation. BC: Berry Curvature. (b) Energy dispersion of the splitting bands, they have Chern number $1$, $-1$, and $1$, respectively.}
\label{figs5}
\end{figure}

Here we show that the $p-q$ zero energy bands can split if the one-body potential lattice is not constructed by the delta potentials.

In realistic experiments, the delta potential can never be realized. Instead, it can be approximated as the linear combination of a series of Fourier components with some truncation,
\begin{equation}
    \hat{V}_{local}=\sum_{i}\sum_{|\bm{q}|\leq{q}_{c}}e^{i\bm{q}\cdot (\hat{\bm{r}}-\bm{r}_i)},
\end{equation}
in which ${q}_{c}$ encloses a circle in crystal momentum space, thus the $\hat{V}_{local}$ is a summation of finite terms. The more terms ${q}_{c}$ includes, the more similar $\hat{V}_{local}$ is to delta potential lattice. When ${q}_{c}$ approaches to infinity, there are infinite terms within this $\hat{V}_{local}$ and it's exactly delta potential lattice. Such truncation makes the dLL split into non-degenerate $p-q$ bands, with energies close to 0. These $p-q$ bands may individually carry Chern numbers different from one, even though their total Chern number is 1. For example, we take $p=4$ and $q=1$, and $\bm{r}_i$ form a square lattice without loss of generality, thus there are three degenerate exact zero energy states when it is a delta potential lattice. However, when the truncation is taken as ${q}_c=2\sqrt{|\bm{b_1}'|^2+|\bm{b_2}'|^2}$, the three zero energy states split and become dispersive, as shown in fig.~\ref{figs5} (b). The Chern numbers of the three bands are $1$, $-1$, and $1$, respectively. The middle band, carrying Chern number $-1$, has heavily non-uniform Berry Curvature, as shown in fig.~\ref{figs5} (a).

\section{ Chern number of decorated LL}

Here we show the Chern number of the decorated LL is also 1. Not losing generality, we take the torus as a square. Taking Landau gauge, the single-body wavefunction on the LLL:
\begin{equation}
    f(x,y)=e^{(-\frac{1}{2}y^2)}\psi(x,y)
\end{equation}
where 
\begin{equation}
    \psi(x,y)=e^{\imath kz}\prod_{\mu=1}^{N_{\Phi}}\theta_1 (\frac{\pi}{L_x}(z-z_{\mu})|\tau).
\end{equation}
The periodic boundary condition asks,
\begin{equation}
    \begin{split}
        & \frac{\psi(z+L_x)}{\psi(z)}=e^{i\theta_x},\\
        & \frac{\psi(z+iL_y)}{\psi(z)}exp \left(i\pi N_{\phi}\left[(2z/L_x)+\tau \right]\right)=e^{i\theta_y}.
    \end{split}
\end{equation}
Therefore,
\begin{equation}
\begin{split}
&k=\frac{2\pi n_1}{L_x}+\frac{\theta_x}{L_x}+\frac{L_y}{2},    \\
& z_0= \frac{N_{\phi}L_x}{2}+\frac{\theta_y L_x}{2\pi}-\tau L_yn_1-\frac{\tau\theta_x L_y}{2\pi}-\frac{\tau N_{\phi}}{2}+n_2 L_x,\\
\end{split}
\end{equation}
where $\theta_x$ and $\theta_y$ are the phases of twisted boundary conditions, and $z_0=\sum z_{\mu}$, $n_1$ and $n_2$ are integers.

As first realized by Thouless and co-workers, a topological property of the wavefunction, known as the first Chern number, can be used to calculate the boundary condition averaged $\sigma_H$, as an integral invariant of many-body wavefunction over twist boundary phase space\cite{niuqian}. And to do calculate the Chern number, we can use the many-body wavefunction, which is a product state of the single particle wavefunction. The wavefunction of Laughlin state is~\cite{Laughlinwavefunction}:
\begin{equation}
    \Psi_0\{z_1, z_2,\cdots\}=e^{-\frac{1}{2}\sum_{i=1}^{N_e}y_i^2}F_{cm}(Z)\prod_{i<j}^{N_e}f(z_i-z_j).
\end{equation}
Above is the wavefunction on the LLL, and on the dLL, which is defined by some delta potentials, the wave function is:
\begin{equation}
    \Psi'\{z_i, \cdots, \eta_j, \cdots \}=e^{-\frac{1}{2}\sum_{i=1}^{N_e}y_i^2}F_{cm}(Z)\prod_{i<j}^{N_e}f(z_i-z_j)\prod_{i=1}^{N_e}\prod_{k=1}^{N_{\delta}}f(z_i-\eta_k).
\end{equation}
in which
\begin{equation}
    f(z_i-z_j)=\theta_1 (\frac{\pi}{L_x}(z_i-z_{j})|\tau),
\end{equation}
\begin{equation}
    f(z_i-\eta_k)=\theta_1 (\frac{\pi}{L_x}(z_i-\eta_k)|\tau),
\end{equation}
\begin{equation}
    F_{cm}(Z)=e^{\imath KZ}\theta_1(\frac{Z-Z_{0}-\sum_k\eta_k}{L_x}|\tau),
\end{equation}
in which $Z=\sum_{i}z_i$. And $\eta_k$ is used to label the position of delta potentials.
In this equation, $N_e=N_{\Phi}-N_{\delta}$ thus the dLL is fulled filled. According to the periodic boundary condition along x direction of centre of mass,
\begin{equation}
    \frac{F_{cm} (Z+L_x),\{z_i\}}{F_{cm} \{z_i\}}=e^{i\theta_x}(-1)^{N_{\phi}-1},
\end{equation}
\begin{equation}
    e^{iKL_x}\frac{\theta_1(\frac{\pi}{L_x}(Z+L_x-Z_0-\sum_k\eta_k)|\tau)}{\theta_1(\frac{\pi}{L_x}(Z-Z_0-\sum_k\eta_k)|\tau)}=e^{i\theta_x} (-1)^{N_{\phi}-1},
\end{equation}
For the Elliptic theta function $\theta_1$, $\frac{\theta_1(z+1)}{\theta_1(z)}=-1$,
thus
\begin{equation}
    e^{iK L_x}=e^{i \theta_x}(-1)^{N_{\phi}}.
\end{equation}
Thus,
\begin{equation}
    KL_x=\theta_x-\pi N_{\phi} -2\pi \cdot b,
\end{equation}
in which $b$ is an integer. Thus,
\begin{equation}
    K=\frac{\theta_x}{L_x}-2\pi \cdot \frac{b}{L_x}-\frac{\pi N_{\phi}}{L_x}
\end{equation}
And similarly, with the periodic boundary condition along y direction:
\begin{equation}
    \frac{F_{cm} (Z+i L_y),\{z_i\}}{F_{cm} \{z_i\}}=e^{i\theta_y} e^{-i\pi \frac{2z}{L_x}+\tau}.
\end{equation}
Also with the quasi-periodicity of theta function, we have:
\begin{equation}
     Z_0=\frac{L_x}{2}(N_{\phi}+\frac{\theta_y}{\pi}+2a)+\imath N_{\phi}K-\sum_i^{N_{\delta}}\eta_i,
\end{equation}
where $a$ and $b$ are integers so that $Z_0$ is in the torus cell $-\frac{L_x}{2}<x\leq \frac{L_x}{2}$, and $-\frac{L_y}{2}<y\leq \frac{L_y}{2}$.

Notice that the $\theta_x$ and $\theta_y$ dependence only appears in the center of mass part of the wavefunction, and the twisting of $\theta_x$ and $\theta_y$ only causes the changing of the phase of many-body wavefunction $\Psi'$, it can be written into,
\begin{equation}
    \Psi'=\Psi e^{\imath \alpha}.
\end{equation}
Notice that $\Psi$ is the wavefunction for periodic boundary condition with $\theta_x=0$ and $\theta_y=0$. Then, 
\begin{equation}
    \begin{split}
        C& = \frac{1}{2\pi i} \oint d\theta_i \bra{\Psi'}\frac{\partial}{\partial \theta_i}\ket{\Psi'}\\
        & = \frac{1}{2\pi i} \oint d\theta_i \bra{\Psi e^{-\imath \alpha} } \frac{\partial}{\partial \theta_i} \ket{\Psi e^{\imath \alpha} }\\
        & = \frac{1}{2\pi i} \oint d\theta_i \bra{\Psi }e^{-\imath \alpha} \frac{\partial}{\partial \theta_i} e^{\imath \alpha} \ket{\Psi}\\
        & = \frac{1}{2\pi i} \oint d\theta_i \iint dx^{Ne}dy^{Ne} |\Psi \{x_{r},y_{r}\}|^2 e^{-\imath \alpha} \frac{\partial}{\partial \theta_i} e^{\imath \alpha}\\
        & = \frac{1}{2\pi i} \oint d\theta_i \iint dx^{Ne}dy^{Ne} |\Psi\{x_{r},y_{r}\}|^2  e^{-\imath \alpha} \cdot \frac{\partial (\imath \alpha)}{\partial \theta_i} \cdot \frac{\partial}{\partial (\imath \alpha)} e^{\imath \alpha}\\
        & = \frac{1}{2\pi i} \oint d\theta_i \iint dx^{Ne}dy^{Ne} |\Psi\{x_{r},y_{r}\}|^2 \frac{\partial (\imath \alpha)}{\partial \theta_i},\\
    \end{split}
\end{equation}
since 
\begin{equation}
    \frac{\partial}{\partial \theta_i}e^{\imath \alpha}=\frac{\partial}{\partial (i\alpha)} \frac{\partial (i\alpha)}{\partial \theta_i}e^{\imath \alpha}.
\end{equation}
where $\frac{\partial}{\partial \theta_i}=\frac{\partial}{\partial \theta_x}+\frac{\partial}{\partial \theta_y}$ is the Nabla operator, and $\{x_r,y_r\}$ expresses the position of $r$-th electron. In our case, the only term which is relevant to $\theta_x$ and $\theta_y$ is the centre of mass part, so we have:
\begin{equation}
\begin{split}
        C & = \frac{1}{2\pi i} \oint d\theta_i \iint dx^{N_e}dy^{N_e} |\Psi\{x_r,y_r\}|^2\frac{\partial}{\partial \theta_i} log(F_{cm}(Z,\theta))\\
        & = \frac{1}{2\pi i} \iint dx^{N_e}dy^{N_e} |\Psi\{x_r,y_r\}|^2 \\
        & \Big\{ \int_0^{2\pi}d\theta_x \frac{\partial}{\partial \theta_i}log(F_{cm}(Z, \theta_x, \theta_y=0))  +\int_0^{2\pi}d\theta_y \frac{\partial}{\partial \theta_i}log(F_{cm}(Z, \theta_x=2\pi, \theta_y))\\
        & +\int_{2\pi}^{0}d\theta_x \frac{\partial}{\partial \theta_i}log(F_{cm}(Z, \theta_x, \theta_y=2\pi)) +\int_{2\pi}^{0}d\theta_y \frac{\partial}{\partial \theta_i}log(F_{cm}(Z, \theta_x=0, \theta_y)) \Big\} \\ 
        & = \frac{1}{2\pi i} \iint dx^{N_e}dy^{N_e} |\Psi\{x_r,y_r\}|^2 \\
        & \Big\{ \int_0^{2\pi}d\theta_x \frac{\partial}{\partial \theta_x} \left [ log(F_{cm}(Z,\theta_x,\theta_y=0))-log(F_{cm}(Z,\theta_x,\theta_y=2\pi)) \right ]\\
        &+\int_0^{2\pi}d\theta_y \frac{\partial}{\partial \theta_y} \left [ log(F_{cm}(Z,\theta_x=2\pi,\theta_y))-log(F_{cm}(Z,\theta_x=0,\theta_y)) \right ] \Big\}\\
        & = \frac{1}{2\pi i} \iint dx^{N_e}dy^{N_e} |\Psi\{x_r,y_r\}|^2 \\
        & \Big\{ \int_0^{2\pi}d\theta_x \frac{\partial}{\partial \theta_x} log \frac{F_{cm}(Z,\theta_x, \theta_y=0)}{F_{cm} (Z,\theta_x,\theta_y=2\pi)}+ \int_0^{2\pi}d\theta_y \frac{\partial}{\partial \theta_y} log \frac{F_{cm}(Z,\theta_x=2\pi,\theta_y)}{F_{cm}(Z,\theta_x=0,\theta_y)}  \Big\}.
\end{split}
\end{equation}
For the convenience of calculation, a constant $M$ is defined such that,
\begin{equation}
    Z_0=\frac{L_x \theta_y}{2\pi}+iN_{\phi}\frac{\theta_x}{L_x}+M.
\end{equation}
Because
\begin{equation}
    \begin{split}
         \frac{\partial}{\partial \theta_y}log\left[ \frac{F_{cm}(Z,\theta_x=2\pi,\theta_y)}{F_{cm}(Z,\theta_x=0,\theta_y)} \right]& =  \frac{\partial}{\partial \theta_y}\bigg\{ log \bigg[ \frac{e^{i K(\theta_x=2\pi)}Z}{e^{i K(\theta_x=0)}Z}\cdot \frac{\theta_1 (\frac{Z-Z_0(\theta_x=2\pi)}{L_x})|\tau)}{\theta_1 (\frac{Z-Z_0(\theta_x=0)}{L_x})|\tau)} \bigg]\bigg \}\\
         & =  \frac{\partial}{\partial \theta_y} \bigg\{ log \bigg[ e^{iZ\frac{2\pi}{L_x}} \frac{\theta_1(\frac{1}{L_x}(Z-M-\frac{L_x \theta_y}{2\pi} -iL_y|\tau)}{\theta_1(\frac{1}{L_x}(Z-M-\frac{L_x \theta_y}{2\pi} -0|\tau)} \bigg] \bigg\}\\
         & =  \frac{\partial}{\partial \theta_y} \bigg\{  log\left[ e^{iZ\frac{2\pi}{lx}} \frac{\theta_1(N-\tau|\tau)}{\theta_1(N|\tau)} \right] \bigg\} \\
         & = \frac{\partial}{\partial \theta_y} \bigg\{ log \left[ e^{iZ\frac{2\pi}{lx}} e^{\pi i (\tau +2N)} \right] \bigg\}\\
         & = \frac{\partial}{\partial \theta_y} \left[ iZ\frac{2\pi}{lx}+  (\pi i (\tau +2N)) \right]\\
         & = i\\
    \end{split}
\end{equation}
where $N$ is defined to satisfy:
\begin{equation}
    N= \frac{1}{L_x}(Z-M-\frac{L_x \theta_y}{2\pi} ).
\end{equation}
We can see the second term of the Chern number expression is obtained by:
\begin{equation}
\begin{split}
     & \frac{1}{2\pi i} \iint dx^{N_e}dy^{N_e} |\Psi\{x_r,y_r\}|^2 \int_0^{2\pi}d\theta_y \frac{\partial}{\partial \theta_y} log \frac{F_{cm}(Z,\theta_x=2\pi,\theta_y)}{F_{cm}(Z,\theta_x=0,\theta_y)} \\
     & = \frac{1}{2\pi i} \iint dx^{N_e}dy^{N_e} |\Psi\{x_r,y_r\}|^2 \int_0^{2\pi}d\theta_y i\\
     & =  \frac{1}{2\pi i} \iint dx^{N_e}dy^{N_e} |\Psi\{x_r,y_r\}|^2 2\pi i\\
     & = 1.\\
\end{split}
\end{equation}
And 
\begin{equation}
    \begin{split}
         & \frac{1}{2\pi i} \iint dx^{N_e}dy^{N_e} |\Psi\{x_r,y_r\}|^2 \int_0^{2\pi}d\theta_x \frac{\partial}{\partial \theta_x} log \frac{F_{cm}(Z,\theta_x,\theta_y=0)}{F_{cm}(Z,\theta_x,\theta_y=2\pi)} \\
         & = \frac{1}{2\pi i} \iint dx^{N_e}dy^{N_e} |\Psi\{x_r,y_r\}|^2 \int_0^{2\pi}d\theta_x \frac{\partial}{\partial \theta_x} log\frac{\theta_1(\frac{1}{L_x}(Z-M-iN_{\phi}\frac{\theta_x}{L_x})|\tau)}{\theta_1(\frac{1}{L_x}(Z-M-L_x-iN_{\phi}\frac{\theta_x}{L_x})|\tau)}\\
         & =  \frac{1}{2\pi i} \iint dx^{N_e}dy^{N_e} |\Psi\{x_r,y_r\}|^2 \int_0^{2\pi}d\theta_x \frac{\partial}{\partial \theta_x} log (e^{i\pi}) \\
         & = 0.
    \end{split}
\end{equation}
 thus
\begin{equation}
    C =1.
\end{equation}

\section{ conductivity of Laughlin states on the decorated LL}

Here we show the zero-energy state of the two-body interaction in null space of delta potentials on the LLL also has Chern number as $1/m$, corresponding to the Laughlin state on the dLL with a filling factor $1/m$, implying the conductivity $\frac{e^2}{mh}$.

With some $\delta$ potentials, the total flux number on the LLL is $N_{\phi}'$, the number of delta potential is $N_{\delta}$, the null space dimension is $N_{\phi}=N_{\phi}'-N_{\delta}$. In order to construct a Laughlin state in the Hilbert space of dLL, the relationship $N_{\phi}=m\cdot N_e$ should be satisfied. And the wavefunction for the Laughlin states is on the dLL:
\begin{equation}
    \Psi_m\{z_1,\dots z_{N_e}\}=e^{-\frac{1}{2}\sum_i^{N_e}y_i^2}F_{cm}\prod_{i<j}^{N_e}f(z_i-z_j)^m\prod_i^{N_e}\prod_{k=1}^{N_{\delta}}f(z_i-\eta_k),
\end{equation}
in which
\begin{equation}
    f(z_i-z_j)=\theta (\frac{\pi}{L_x}(z_i-z_{j})|\tau),
\end{equation}
\begin{equation}
    f(z_i-\eta_k)=\theta (\frac{\pi}{L_x}(z_i-\eta_k)|\tau),
\end{equation}
\begin{equation}
    F_{cm}(Z)=e^{\imath KZ}\prod_{\nu}^{m}\theta_1(\frac{Z-Z_{\nu}-\frac{\sum_k\eta_k}{m}}{L_x}|\tau),
\end{equation}
in which $Z=\sum_{i}z_i$. And $\eta_k$ is used to label the position of delta potentials. Similar as the product state on the dLL, the $\theta_x$ and $\theta_y$ dependence only appears in the center of mass part of the wavefunction of Laughlin state on the dLL, and the twisting of boundary conditions only causes the changing of the phase of many-body wavefunction. Then the wavefunction can be written into,
\begin{equation}
    \Psi_m'=\Psi_me^{i \alpha}.
\end{equation}
However, different from the case of IQHE state, the range for the Laughlin states is $[0,m\cdot 2\pi]$ since one of the states gets back to itself only after passing through the twisted angle $m\cdot 2\pi$. The Chern number can be calculated by:
\begin{equation}
      C = \frac{1}{2\pi i m^2} \oint d\theta_i \iint dx^{N_e}dy^{N_e} |\Psi_m\{x_r,y_r\}|^2\frac{\partial}{\partial \theta_i} log(F_{cm}(Z,\theta)).
\end{equation}

Thus,
\begin{equation}
\begin{split}
        C =  \frac{1}{2\pi i m^2} \iint dx^{N_e}dy^{N_e} |\Psi_m\{x_r,y_r\}|^2 &\bigg\{ \int_0^{m 2\pi}d\theta_x \frac{\partial}{\partial \theta_x} log \frac{F_{cm}(Z,\theta_x, \theta_y=0)}{F_{cm} (Z,\theta_x,\theta_y=2\pi\cdot m)}\\ 
        & + \int_0^{m 2\pi}d\theta_y \frac{\partial}{\partial \theta_y} log \frac{F_{cm}(Z,\theta_x=2\pi\cdot m,\theta_y)}{F_{cm}(Z,\theta_x=0,\theta_y)} \bigg\}.
\end{split}
\end{equation}
The periodicity of centre of mass gives:
\begin{equation}
    \frac{F_{cm}(Z+L_x),\{z_r\}}{F_{cm}(Z),\{z_r\}}=(-1)^{N_{\phi}-m}e^{i\theta_x}.
\end{equation}
Thus,
\begin{equation}
    e^{iKL_x}=(-1)^{N_{\phi}}e^{i\theta_x}.
\end{equation}
thus,
\begin{equation}
    KL_x=\theta_x-2\pi b -\pi N_{\phi},
\end{equation}
in which $b$ is an integer.
\begin{equation}
    K=\frac{\theta_x}{L_x}-2\pi\cdot \frac{b}{L_x}-\frac{\pi N_{\phi}}{L_x}.
\end{equation}
Similarly, 
\begin{equation}
    \frac{F_{cm}(Z+i L_y),\{z_r\}}{F_{cm}(Z),\{z_r\}}\cdot e^{i\pi m( \frac{2Z}{L_x}+\tau)}=(-1)^{N_{\phi}-m}e^{i\theta_y}.
\end{equation}
thus,
\begin{equation}
     \sum_{\nu}Z_{\nu}=\frac{L_x}{2}(N_{\phi}+\frac{\theta_y}{\pi}+2a)+\imath N_{\phi}K-\sum_i^{N_{\delta}}\eta_i.
\end{equation}
saying $Z_s=\sum_{\nu}Z_{\nu}$, and the average of $Z_s$ is:
\begin{equation}
    Z_a=\frac{Z_s}{m}
\end{equation}
that is:
\begin{equation}
    Z_a=\frac{N_{_{\phi}}L_x}{2m}+\frac{a L_x}{m}+\frac{L_x \theta_y}{2\pi m}-i\frac{K N_{\phi}}{m}-\frac{\sum_{k}^{N_{\delta}}\eta_k}{m}
\end{equation}
Let suppose, for the $\nu-th$ zero point of the centre of mass part, we have:
\begin{equation}
    Z_{\nu}=Z_a+d_{\nu},
\end{equation}
where $d_{\nu}$ is the distance between the average zero point and the real zero point. Notice that twisted angle only pushed the position of average zero point to move but don't change the relative position of these zero points of centre of mass part, which means $d_{\nu}$ is constant.

\begin{equation}
    Z_{\nu}=\frac{N_{_{\phi}}L_x}{2m}+\frac{a L_x}{m}+\frac{L_x \theta_y}{2\pi m}-i\frac{K N_{\phi}}{m}-\frac{\sum_{k}^{N_{\delta}}\eta_k}{m}+d_{\nu}.
\end{equation}
Since 
\begin{equation}
    K=\frac{\theta_x}{L_x}-2\pi\cdot \frac{b}{L_x}-\frac{\pi N_{\phi}}{L_x},
\end{equation}
m constants $P_{\nu}$ are defined, such that,
\begin{equation}
    Z_{\nu}=\frac{L_x\theta_y}{2\pi\cdot m}+iN_{\phi}\frac{\theta_x}{mL_x}+P_{\nu}.
\end{equation}
Therefore,
\begin{equation}
    \begin{split}
         & \frac{1}{2\pi i m^2} \iint dx^{N_e}dy^{N_e} |\Psi\{x_r,y_r\}|^2 \int_0^{2\pi\cdot m}d\theta_x \frac{\partial}{\partial \theta_x} log \frac{F_{cm}(Z,\theta_x,\theta_y=0)}{F_{cm}(Z,\theta_x,\theta_y=2\pi\cdot m)} \\
         & = \frac{1}{2\pi i m^2} \iint dx^{N_e}dy^{N_e} |\Psi\{x_r,y_r\}|^2 \int_0^{2\pi\cdot m}d\theta_x \frac{\partial}{\partial \theta_x} log\bigg[\prod_{\nu}^m \frac{\theta_1(\frac{1}{L_x}(Z-P_{\nu}-iN_{\phi}\frac{\theta_x}{L_x})|\tau)}{\theta_1(\frac{1}{L_x}(Z-P_{\nu}-L_x-iN_{\phi}\frac{\theta_x}{L_x})|\tau)}\bigg]\\
         & =  \frac{1}{2\pi i m^2} \iint dx^{N_e}dy^{N_e} |\Psi\{x_r,y_r\}|^2 \int_0^{2\pi}d\theta_x \frac{\partial}{\partial \theta_x} log (e^{i\pi})^3 \\
         & = 0.
    \end{split}
\end{equation}
\begin{equation}
\begin{split}
     & \frac{1}{2\pi i m^2} \iint dx^{N_e}dy^{N_e} |\Psi\{x_r,y_r\}|^2 \int_0^{2\pi\cdot m}d\theta_y \frac{\partial}{\partial \theta_y} log \frac{F_{cm}(Z,\theta_x=2\pi\cdot m,\theta_y)}{F_{cm}(Z,\theta_x=0,\theta_y)} \\
     & = \frac{1}{2\pi i m^2} \iint dx^{N_e}dy^{N_e} |\Psi\{x_r,y_r\}|^2 \int_0^{2\pi\cdot m}d\theta_y m\cdot i\\
     & =  \frac{1}{2\pi i m^2} \iint dx^{N_e}dy^{N_e} |\Psi\{x_r,y_r\}|^2 2\pi i \cdot m\\
     & = \frac{1}{m}.\\
\end{split}
\end{equation}
Thus,
\begin{equation}
    C=0+\frac{1}{m}=\frac{1}{m}.
\end{equation}

\section{ Occupation schematics of phases and conductivity calculation}

Here we show different phases with the competition between the local potentials and interactions and various filling factors, and we also give the numerical method of calculating the conductivity.

\begin{figure}[ht]
\centering
\includegraphics[width=0.5 \linewidth]{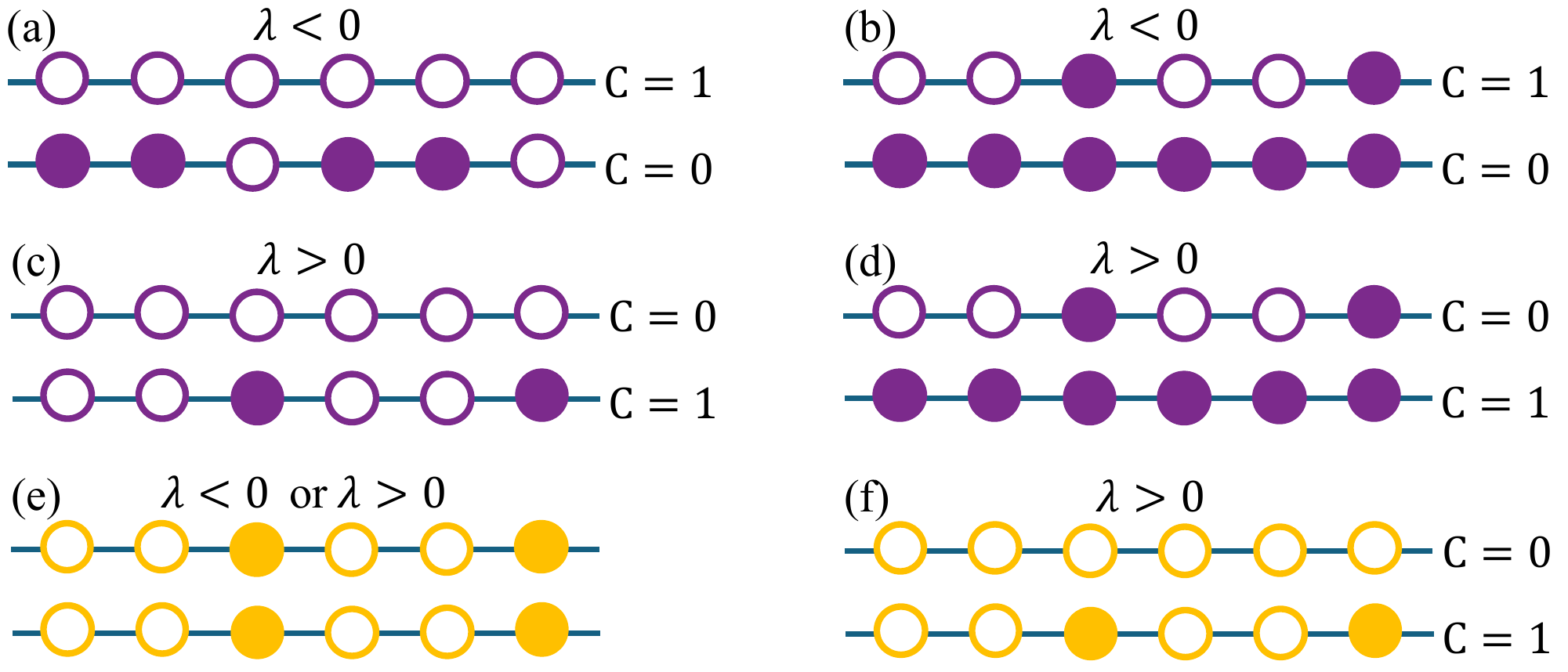}
\caption{\textbf{Schematics of occupation on the dLL and the dispersive bands in the presence of $V_{int}$ and delta potential lattice.} Panels (a)–(d) correspond to the regime of interaction as a perturbation, whereas panels (e) and (f) correspond to the regime of dominating interaction. (a) Electrons are all trapped within the dispersive bands due to negative large local potentials, leading to Fermi liquid with a non-quantized conductivity. (b) Negative large local potentials make the dispersive bands fully filled and the dLL is one-third filled, resulting into $\sigma_{xy}=1/3$ while the total filling factor of the LLL is $2/3$. (c) The dLL is one-third filled while the dispersive band is empty due to large positive delta potentials, band mixing is suppressed, resulting a gapped topological phase with conductivity $1/3$. (d) The dLL is fully filled while the dispersive bands are partially filled, leading to the Fermi liquid with a non-quantized conductivity. (e) Strong interactions causes band mixing between the dLL and dispersive bands, and the gapped topological phase is robust when $\nu_l=1/3$, offering $\sigma_{xy}=1/3$. (f) With positive delta potentials, band mixing between the dLL and dispersive bands is suppressed with short range interactions. The dLL is partially filled and the dispersive bands are empty. It is a gapped topological phase with gap scale of local potentials, giving $\sigma_{xy}=1/3$.
}
\label{fig2}
\end{figure}

As shown in fig.~\ref{fig2}, in both the interaction-dominated and local-potential-dominated regimes, the system may give a gapped topological phase or Fermi liquid as filling factor changes. And more interesting may appear when $\lambda$ changes, as we show in the phase diagram in the main text.

In the study of quantum Hall effects and topological states of matter, the quantization of Hall conductivity is intimately connected to the topological properties of the quantum wave function. As first established by Thouless and coworkers in the theory of topological quantum transport, the Hall conductivity can be expressed in terms of a global topological invariant of the many‑body ground state—the  Chern number~\cite{niuqian}. 

For a two‑dimensional periodic system, one introduces twist boundary phases $\theta_x$ and $\theta_y$ to parameterize different boundary conditions. The Chern number is defined as an integral of the Berry curvature over the torus of boundary phases:
\begin{equation}
    C = \frac{1}{2\pi} \int_{0}^{2\pi} \! \int_{0}^{2\pi} 
\mathcal{F}(\theta_x,\theta_y) \; d\theta_x d\theta_y,
\end{equation}
where the Berry curvature $\mathcal{F}(\theta_x,\theta_y)$ is given by
\begin{equation}
    \mathcal{F}(\theta_x,\theta_y) = 
\partial_{\theta_x} \mathcal{A}_2(\theta_x,\theta_y) 
- \partial_{\theta_y} \mathcal{A}_1(\theta_x,\theta_y),
\end{equation}
with $\mathcal{A}_\mu(\theta_x,\theta_y) = i \langle \psi(\theta_x,\theta_y) | 
\partial_{\theta_\mu} | \psi(\theta_x,\theta_y) \rangle$ being the Berry connection.

To evaluate the Chern number numerically, the phase space $[0,2\pi]\times[0,2\pi]$ is discretized into a uniform mesh, and the Chern number is thus approximately given by the summation of Berry phase of each plaquette\cite{DNsheng}. A common choice is to take $20\times 20$ grid points, i.e. 
\[
\delta\theta = \frac{2\pi}{20},
\qquad 
\theta_x^{(m)} = m\,\delta\theta,\quad 
\theta_y^{(n)} = n\,\delta\theta,\quad m,n=0,1,\dots,19 .
\]
Each elementary plaquette is a square of side $\delta\theta$ with four vertices (taken in counter‑clockwise order):
\begin{align*}
& P_1: (\theta_x,\theta_y), \\
& P_2: (\theta_x+\delta\theta,\theta_y), \\
& P_3: (\theta_x+\delta\theta,\theta_y+\delta\theta), \\
& P_4: (\theta_x,\theta_y+\delta\theta) .
\end{align*}
Let $|\psi_i\rangle$ ($i=1,2,3,4$) be the normalized ground‑state wave function at the corresponding twist angles.

The Berry phase (or the gauge flux) associated with the plaquette is defined by the cyclic product of inner products:
\begin{equation}
    \Omega(\theta_x,\theta_y) 
= \langle \psi_1|\psi_2\rangle \,
  \langle \psi_2|\psi_3\rangle \,
  \langle \psi_3|\psi_4\rangle \,
  \langle \psi_4|\psi_1\rangle .
\end{equation}
For a sufficiently fine mesh the phase of $\Omega$ is small; its imaginary part gives the Berry flux through the plaquette:
\begin{equation}
    \Phi_{\text{B}}(\theta_x,\theta_y) = \operatorname{Im}\ln \Omega(\theta_x,\theta_y).
\end{equation}

Summing the fluxes over all plaquettes yields the discretized Chern number:
\begin{equation}
    C = \frac{1}{2\pi} \sum_{\theta_x,\theta_y} 
\Phi_{\text{B}}(\theta_x,\theta_y).
\end{equation}

Because the Chern number is a topological integer, the result should be very close to an integer (up to numerical errors). In a quantum Hall system this integer directly gives the Hall conductivity in units of the conductance quantum $e^2/h$:
\begin{equation}
    \sigma_H = C \, \frac{e^2}{h}.
\end{equation}

\section{ Equivalence of dLL and LLL from projected Interactions}

Here we show that the interaction projected on the dLL can be understood as an effective interaction projected onto the LLL~\cite{Equivalence}. 

On the LLL, any density-density interaction can be written into,
\begin{equation}
    \hat{V}=\sum_{\bm{k}_1,\bm{k}_2,\bm{k}_3,\bm{k}_4}V_{\bm{k}_1,\bm{k}_2,\bm{k}_3,\bm{k}_4}c^{\dagger}_{\bm{k}_1}c^{\dagger}_{\bm{k}_2}c_{\bm{k}_3}c_{\bm{k}_4}\delta'(\bm{k}_1+\bm{k}_2,\bm{k}_3+\bm{k}_4),
\end{equation}
in which 
\begin{equation}
\begin{split}
    V_{\bm{k}_1,\bm{k}_2,\bm{k}_3,\bm{k}_4} & =\bra{\bm{k}_1,\bm{k}_2}\hat{V}\ket{\bm{k}_3,\bm{k}_4}\\
    & = \bra{\bm{k}_1,\bm{k}_2}\sum_{\bm{q}}V(\bm{q})e^{i\bm{q}\cdot \bm{r}}\ket{\bm{k}_3,\bm{k}_4}\\
    & = \sum_{\bm{q}}V(\bm{q})e^{-\frac{1}{2}|\bm{q}|^2}\bra{\bm{k}_1}e^{i\bm{q}\cdot \bar{\bm{R}}}\ket{\bm{k}_4} \bra{\bm{k}_3}e^{-i\bm{q}\cdot \bar{\bm{R}}}\ket{\bm{k}_4}.\\
\end{split}
\end{equation}
$\delta'(\bm{k}_1+\bm{k}_2,\bm{k}_3+\bm{k}_4)$ means $\bm{k}_1+\bm{k}_2$ and $\bm{k}_3+\bm{k}_4$ differs at most a reciprocal lattice vector.
Notice the periodic boundary condition gives:
\begin{equation}
    \ket{\bm{k}+\bm{b}}=\eta_{\bm{b}} e^{-\frac{i}{2}\bm{b}\times \bm{k}}\ket{\bm{k}},
\end{equation}
in which $\eta_{\bm{b}}=1$ if ${\bm{b}}/2$ is reciprocal lattice and  $\eta_{\bm{b}}=-1$ if ${\bm{b}}/2$ is not.
\begin{equation}
    e^{i\bm{q}\hat{\bar{\bm{R}}}}\ket{\bm{k}}=e^{\frac{i}{2}\bm{q}\times \bm{k}}\ket{\bm{k}+\bm{q}}.
\end{equation}
\begin{equation}
    \begin{split}
        & \bra{\bm{k}}e^{i\bm{q}\hat{\bar{\bm{R}}}}\ket{\bm{k}'} \\ 
        &=\bra{k\bm{k}}e^{\frac{i}{2}\bm{q}\times \bm{k}'}\ket{\bm{k}'+\bm{q}}\\
        & \propto \delta'(\bm{k},\bm{k}'+\bm{q})\\
    \end{split}
\end{equation}
Saying $\bm{k}=\bm{k}'+\bm{q}+\bm{b}$, that is:
\begin{equation}
    \bm{q}=\bm{k}-\bm{k}'-\bm{b}.
\end{equation}
Then we have:
\begin{equation}
    \begin{split}
        \bra{\bm{k}}e^{i\bm{q}\hat{\bar{\bm{R}}}}\ket{\bm{k}'} & = e^{\frac{i}{2}\bm{q}\times \bm{k}'}\bra{\bm{k}} \eta_{\bm{b}} e^{\frac{i}{2}\bm{b}\times (\bm{k}'+\bm{q})}\ket{\bm{k}'+\bm{q}+\bm{b}}\\
        & = e^{\frac{i}{2}\bm{q}\times \bm{k}'} \eta_{\bm{b}} e^{\frac{i}{2}\bm{b}\times (\bm{k}'+\bm{q})} \delta(\bm{q},\bm{k}-\bm{k}'-\bm{b})\\
        & = \eta_{\bm{b}} e^{\frac{i}{2}(\bm{k}-\bm{k}'-\bm{b})\times \bm{k}'}e^{\frac{i}{2}\bm{b}\times (\bm{k}-\bm{b})}\\
        & = \eta_{\bm{b}}e^{\frac{i}{2}\bm{k}\times \bm{k}'}e^{\frac{i}{2}(\bm{k}+\bm{k}')\times \bm{b}}.
    \end{split}
\end{equation}
Thus, 
\begin{equation}
     \begin{split}
         V_{\bm{k}_1,\bm{k}_2,\bm{k}_3,\bm{k}_4}& =\sum_{\bm{q}}V(\bm{q})e^{-\frac{1}{2}|\bm{q}|^2}\eta_{\bm{b}}e^{\frac{i}{2}\bm{k}_1\times \bm{k}_4}e^{\frac{i}{2}(\bm{k}_1+\bm{k}_4)\times \bm{b}}\eta_{-\bm{b}+\Delta \bm{b}}e^{\frac{i}{2}\bm{k}_2\times \bm{k}_3}e^{\frac{i}{2}(\bm{k}_2+\bm{k}_3)\times (-\bm{b}+\Delta \bm{b})}\\
         & = \sum_{\bm{b}}V(\bm{k}_1-\bm{k}_4-\bm{b}) f^{\bm{k}_1,\bm{k}_4}_{\bm{b}}f^{\bm{k}_2,\bm{k}_3}_{-\bm{b}+\Delta \bm{b}}\\
         & = \sum_{\bm{b}}V(\bm{k}_1-\bm{k}_4-\bm{b}) F^{\bm{b},\Delta\bm{b}}_{\bm{k}_1,\bm{k}_2,\bm{k}_3,\bm{k}_4},\\
     \end{split}
\end{equation}
in which 
\begin{equation}
    f_{\bm{b}}^{\bm{k},\bm{k}'}=e^{-\frac{|\bm{k}-\bm{k}'-\bm{b}|^2}{4}}\eta_{\bm{b}}e^{\frac{i}{2}\bm{k}\times \bm{k}'}e^{\frac{i}{2}(\bm{k}+\bm{k}')\times \bm{b}},
\end{equation}
\begin{equation}
    \Delta \bm{b}=\bm{k}_1+\bm{k}_2-\bm{k}_3-\bm{k}_4.
\end{equation}
\begin{equation}
    F^{\bm{b},\Delta\bm{b}}_{\bm{k}_1,\bm{k}_2,\bm{k}_3,\bm{k}_4}=f_{\bm{b}}^{\bm{k}_1,\bm{k}_4}\cdot f_{-\bm{b}+\Delta\bm{b}}^{\bm{k}_2,\bm{k}_3}.
\end{equation}
Similarly, any density-density interaction on the dLL can also be written into,
\begin{equation}
    \hat{V}=\sum_{\bm{k}_1,\bm{k}_2,\bm{k}_3,\bm{k}_4}U_{\bm{k}_1,\bm{k}_2,\bm{k}_3,\bm{k}_4}c^{\dagger}_{\bm{k}_1}c^{\dagger}_{\bm{k}_2}c_{\bm{k}_3}c_{\bm{k}_4}\delta'(\bm{k}_1+\bm{k}_2,\bm{k}_3+\bm{k}_4),
\end{equation}
in which 
\begin{equation}
\begin{split}
    U_{\bm{k}_1,\bm{k}_2,\bm{k}_3,\bm{k}_4} & =\bra{\phi_{\bm{k}_1},\phi_{\bm{k}_2}}\hat{V}\ket{\phi_{\bm{k}_3},\phi_{\bm{k}_4}}\\
    & = \sum_{\bm{q}}V(\bm{q})e^{-\frac{1}{2}|\bm{q}|^2}\bra{\phi_{\bm{k}_1}}e^{i\bm{q}\cdot \bar{\bm{R}}}\ket{\phi_{\bm{k}_4}} \bra{\phi_{\bm{k}_3}}e^{-i\bm{q}\cdot \bar{\bm{R}}}\ket{\phi_{\bm{k}_4}},\\
\end{split}
\end{equation}
in which $\ket{\phi_{\bm{k}}}$ is the Bloch basis on the dLL, according to the previous section, 
\begin{equation}
    \phi_{\bm{k}}(\bm{r})=c_{\bm{k}}\psi_{\bm{k}}(\bm{r})+c_{\bm{k}+\bm{b}_2}\psi_{\bm{k}+\bm{b}_2}(\bm{r}).
\end{equation}
Thus, 
\begin{equation}
\begin{split}
    \bra{\phi_{\bm{k}}}e^{i\bm{q}\cdot \bar{\bm{R}}}\ket{\phi_{\bm{k}'}} &=c_{\bm{k}}^*c_{\bm{k}'}\bra{\bm{k}}e^{i\bm{q}\hat{\bar{\bm{R}}}}\ket{\bm{k}'}+c_{\bm{k}+\bm{b}_2}^*c_{\bm{k}'+\bm{b}_2}\bra{\bm{k}+\bm{b}_2}e^{i\bm{q}\hat{\bar{\bm{R}}}}\ket{\bm{k}'+\bm{b}_2}\\
    &= c_{\bm{k}}^*c_{\bm{k}'}f_{\bm{b}}^{\bm{k},\bm{k}'}+c_{\bm{k}+\bm{b}_2}^*c_{\bm{k}'+\bm{b}_2}f_{\bm{b}}^{\bm{k}+\bm{b}_2,\bm{k}'+\bm{b}_2},\\
\end{split}
\end{equation}
in which $\bm{q}=\bm{k}-\bm{k}'-\bm{b}$. $\bm{b}_1$ and $\bm{b}_2$ are the reciprocal lattice vectors of the lattice potential, $\bm{b}_1'$ and $\bm{b}_2'$ can be constructed as the reciprocal lattice vectors of the magnetic unit, by $\bm{b}_1'=\bm{b}_1$, and $\bm{b}_2'=2\bm{b}_2$ without loss of generality. Then the matrix elements of the two-body interaction on the dLL can be written into,
\begin{equation}
    \begin{split}
        U_{\bm{k}_1,\bm{k}_2,\bm{k}_3,\bm{k}_4} &= \sum_{\bm{q}}V(\bm{q})e^{-\frac{1}{2}|\bm{q}|^2}\bra{\phi_{\bm{k}_1}}e^{i\bm{q}\cdot \bar{\bm{R}}}\ket{\phi_{\bm{k}_4}} \bra{\phi_{\bm{k}_2}}e^{-i\bm{q}\cdot \bar{\bm{R}}}\ket{\phi_{\bm{k}_3}}\\
        & = \sum_{\bm{b}}V(\bm{k}_1-\bm{k}_4-\bm{b}) \cdot \\
        & \left[ c_{\bm{k}_1}^*c_{\bm{k}_4}f_{\bm{b}}^{\bm{k}_1,\bm{k}_4}+c_{\bm{k}_1+\bm{b}_2}^*c_{\bm{k}_4+\bm{b}_2}f_{\bm{b}}^{\bm{k}_1+\bm{b}_2,\bm{k}_4+\bm{b}_2}\right] \cdot \left[c_{\bm{k}_2}^*c_{\bm{k}_3}f_{-\bm{b}+\Delta\bm{b}}^{\bm{k}_2,\bm{k}_3}+c_{\bm{k}_2+\bm{b}_2}^*c_{\bm{k}_3+\bm{b}_2}f_{-\bm{b}+\Delta\bm{b}}^{\bm{k}_2+\bm{b}_2,\bm{k}_3+\bm{b}_2}\right]\\
        &  = \sum_{\bm{b}}V(\bm{k}_1-\bm{k}_4-\bm{b}) \cdot \bigg[c_{\bm{k}_1}^*c_{\bm{k}_4}c_{\bm{k}_2}^*c_{\bm{k}_3} f_{\bm{b}}^{\bm{k}_1,\bm{k}_4} \cdot f_{-\bm{b}+\Delta\bm{b}}^{\bm{k}_2,\bm{k}_3} + c_{\bm{k}_1+\bm{b}_2}^*c_{\bm{k}_4+\bm{b}_2}c_{\bm{k}_2}^*c_{\bm{k}_3} f_{\bm{b}}^{\bm{k}_1+\bm{b}_2,\bm{k}_4+\bm{b}_2} \cdot f_{-\bm{b}+\Delta\bm{b}}^{\bm{k}_2,\bm{k}_3}\\
        & +c_{\bm{k}_1}^*c_{\bm{k}_4}c_{\bm{k}_2+\bm{b}_2}^*c_{\bm{k}_3+\bm{b}_2} f_{\bm{b}}^{\bm{k}_1,\bm{k}_4} \cdot f_{-\bm{b}+\Delta\bm{b}}^{\bm{k}_2+\bm{b}_2,\bm{k}_3+\bm{b}_2} +c_{\bm{k}_1+\bm{b}_2}^*c_{\bm{k}_4+\bm{b}_2}c_{\bm{k}_2+\bm{b}_2}^*c_{\bm{k}_3+\bm{b}_2} f_{\bm{b}}^{\bm{k}_1+\bm{b}_2,\bm{k}_4+\bm{b}_2} \cdot f_{-\bm{b}+\Delta\bm{b}}^{\bm{k}_2+\bm{b}_2,\bm{k}_3+\bm{b}_2} \bigg]\\
        & = \sum_{\bm{b}}V(\bm{k}_1-\bm{k}_4-\bm{b}) \cdot \bigg[c_{\bm{k}_1}^*c_{\bm{k}_4}c_{\bm{k}_2}^*c_{\bm{k}_3}F^{\bm{b},\Delta\bm{b}}_{\bm{k}_1,\bm{k}_2,\bm{k}_3,\bm{k}_4}+c_{\bm{k}_1+\bm{b}_2}^*c_{\bm{k}_4+\bm{b}_2}c_{\bm{k}_2}^*c_{\bm{k}_3}F^{\bm{b},\Delta\bm{b}}_{\bm{k}_1+\bm{b}_2,\bm{k}_2,\bm{k}_3,\bm{k}_4+\bm{b}_2}\\
        & +c_{\bm{k}_1}^*c_{\bm{k}_4}c_{\bm{k}_2+\bm{b}_2}^*c_{\bm{k}_3+\bm{b}_2}F^{\bm{b},\Delta\bm{b}}_{\bm{k}_1,\bm{k}_2+\bm{b}_2,\bm{k}_3+\bm{b}_2,\bm{k}_4}+c_{\bm{k}_1+\bm{b}_2}^*c_{\bm{k}_4+\bm{b}_2}c_{\bm{k}_2+\bm{b}_2}^*c_{\bm{k}_3+\bm{b}_2} F^{\bm{b},\Delta\bm{b}}_{\bm{k}_1+\bm{b}_2,\bm{k}_2+\bm{b}_2,\bm{k}_3+\bm{b}_2,\bm{k}_4+\bm{b}_2} \bigg].\\
    \end{split}
\end{equation}
The two-body interaction matrix element $U_{\bm{k}_1,\bm{k}_2,\bm{k}_3,\bm{k}_4}$ on the dLL can also be written into this form:
\begin{equation}
    U_{\bm{k}_1,\bm{k}_2,\bm{k}_3,\bm{k}_4}= \sum_{\bm{b}}V(\bm{k}_1-\bm{k}_4-\bm{b}) \cdot G^{\bm{b},\Delta\bm{b}}_{\bm{k}_1,\bm{k}_2,\bm{k}_3,\bm{k}_4}.
\end{equation}
Thus, 
\begin{equation}
    \begin{split}
        G^{\bm{b},\Delta\bm{b}}_{\bm{k}_1,\bm{k}_2,\bm{k}_3,\bm{k}_4}& = c_{\bm{k}_1}^*c_{\bm{k}_4}c_{\bm{k}_2}^*c_{\bm{k}_3}F^{\bm{b},\Delta\bm{b}}_{\bm{k}_1,\bm{k}_2,\bm{k}_3,\bm{k}_4}+c_{\bm{k}_1+\bm{b}_2}^*c_{\bm{k}_4+\bm{b}_2}c_{\bm{k}_2}^*c_{\bm{k}_3}F^{\bm{b},\Delta\bm{b}}_{\bm{k}_1+\bm{b}_2,\bm{k}_2,\bm{k}_3,\bm{k}_4+\bm{b}_2}\\
        & +c_{\bm{k}_1}^*c_{\bm{k}_4}c_{\bm{k}_2+\bm{b}_2}^*c_{\bm{k}_3+\bm{b}_2}F^{\bm{b},\Delta\bm{b}}_{\bm{k}_1,\bm{k}_2+\bm{b}_2,\bm{k}_3+\bm{b}_2,\bm{k}_4}+c_{\bm{k}_1+\bm{b}_2}^*c_{\bm{k}_4+\bm{b}_2}c_{\bm{k}_2+\bm{b}_2}^*c_{\bm{k}_3+\bm{b}_2} F^{\bm{b},\Delta\bm{b}}_{\bm{k}_1+\bm{b}_2,\bm{k}_2+\bm{b}_2,\bm{k}_3+\bm{b}_2,\bm{k}_4+\bm{b}_2}\\
        & = F^{\bm{b},\Delta\bm{b}}_{\bm{k}_1,\bm{k}_2,\bm{k}_3,\bm{k}_4} \cdot \bigg[c_{\bm{k}_1}^*c_{\bm{k}_4}c_{\bm{k}_2}^*c_{\bm{k}_3}+c_{\bm{k}_1+\bm{b}_2}^*c_{\bm{k}_4+\bm{b}_2}c_{\bm{k}_2}^*c_{\bm{k}_3}\frac{F^{\bm{b},\Delta\bm{b}}_{\bm{k}_1+\bm{b}_2,\bm{k}_2,\bm{k}_3,\bm{k}_4+\bm{b}_2}}{F^{\bm{b},\Delta\bm{b}}_{\bm{k}_1,\bm{k}_2,\bm{k}_3,\bm{k}_4}} \\
        &+c_{\bm{k}_1}^*c_{\bm{k}_4}c_{\bm{k}_2+\bm{b}_2}^*c_{\bm{k}_3+\bm{b}_2}\frac{F^{\bm{b},\Delta\bm{b}}_{\bm{k}_1,\bm{k}_2+\bm{b}_2,\bm{k}_3+\bm{b}_2,\bm{k}_4}}{F^{\bm{b},\Delta\bm{b}}_{\bm{k}_1,\bm{k}_2,\bm{k}_3,\bm{k}_4}}+ c_{\bm{k}_1+\bm{b}_2}^*c_{\bm{k}_4+\bm{b}_2}c_{\bm{k}_2+\bm{b}_2}^*c_{\bm{k}_3+\bm{b}_2} \frac{F^{\bm{b},\Delta\bm{b}}_{\bm{k}_1+\bm{b}_2,\bm{k}_2+\bm{b}_2,\bm{k}_3+\bm{b}_2,\bm{k}_4+\bm{b}_2}}{F^{\bm{b},\Delta\bm{b}}_{\bm{k}_1,\bm{k}_2,\bm{k}_3,\bm{k}_4}}
        \bigg]\\
        & = F^{\bm{b},\Delta\bm{b}}_{\bm{k}_1,\bm{k}_2,\bm{k}_3,\bm{k}_4} \cdot \bigg[c_{\bm{k}_1}^*c_{\bm{k}_4}c_{\bm{k}_2}^*c_{\bm{k}_3}+c_{\bm{k}_1+\bm{b}_2}^*c_{\bm{k}_4+\bm{b}_2}c_{\bm{k}_2}^*c_{\bm{k}_3}\cdot e^{\frac{i}{2}\bm{b}_2\times (\bm{k}_4-\bm{k}_1)}\\
        & + c_{\bm{k}_1}^*c_{\bm{k}_4}c_{\bm{k}_2+\bm{b}_2}^*c_{\bm{k}_3+\bm{b}_2}\cdot e^{\frac{i}{2}\bm{b}_2\times (\bm{k}_3-\bm{k}_2)} + c_{\bm{k}_1+\bm{b}_2}^*c_{\bm{k}_4+\bm{b}_2}c_{\bm{k}_2+\bm{b}_2}^*c_{\bm{k}_3+\bm{b}_2} \cdot e^{\frac{i}{2}\bm{b}_2\times (-\Delta\bm{b})} \bigg]\\
    \end{split}
\end{equation}
Since $\bm{q}=\bm{k}_1-\bm{k}_4-\bm{b}$, one can have $\bm{k}_4=\bm{k}_1-\bm{q}-\bm{b}$.  Because the range of $\bm{k}_4$ is in the first Brillouin Zone and $\bm{b}=m\bm{b}_1+n\bm{b}_2$, $\bm{b}$ is unique to make sure $\bm{k}_4$ is in the first Brillouin Zone. Thus $F^{\bm{b},\Delta\bm{b}}_{\bm{k}_1,\bm{k}_2,\bm{k}_3,\bm{k}_4}$ only depends on $\bm{k}_1$, $\bm{k}_2$ and $\bm{q}$, so as $G^{\bm{b},\Delta\bm{b}}_{\bm{k}_1,\bm{k}_2,\bm{k}_3,\bm{k}_4}$. One can write $\bm{k}_4$ as:
\begin{equation}
    \bm{k}_4=Mod(\bm{k}_1-\bm{q}),
\end{equation}
in which $Mod(\bm{p})$ is defined as the modulo function, $Mod(\bm{p})=\bm{b}-m\bm{b}_1-n\bm{b}_2=p^x\bm{b}_1+p^y\bm{b}_2$, where integers $m$ and $n$ are chose such that $p^{x(y)}\in [0,1)$. Thus
\begin{equation}
    \bm{b}=\bm{k}_1-\bm{k}_4-\bm{q}=\bm{k}_1-\bm{q}-Mod(\bm{k}_1-\bm{q}).
\end{equation}
Similarly,
\begin{equation}
    \bm{k}_3=Mod(\bm{k}_2+\bm{q}).
\end{equation}
Thus,
\begin{equation}
    \begin{split}
        F_{\bm{k}_1,\bm{k}_2}^{\bm{q}}&=\eta_{\bm{k}_1-\bm{q}-Mod(\bm{k}_1-\bm{q})}e^{\frac{i}{2}(\bm{k}_1+Mod(\bm{k}_1-\bm{q}))\times (\bm{k}_1-\bm{q}-Mod(\bm{k}_1-\bm{q}))}e^{\frac{i}{2}\bm{k}_1\times Mod(\bm{k}_1-\bm{q})}e^{-\frac{1}{4}|\bm{q}|^2}  \\
        & = \eta_{\bm{k}_2+\bm{q}-Mod(\bm{k}_2+\bm{q})}e^{\frac{i}{2}(\bm{k}_2+Mod(\bm{k}_2+\bm{q}))\times (\bm{k}_2+\bm{q}-Mod(\bm{k}_2+\bm{q}))}e^{\frac{i}{2}\bm{k}_2\times Mod(\bm{k}_2+\bm{q})}e^{-\frac{1}{4}|\bm{q}|^2}
        \end{split}
\end{equation}
And,
\begin{equation}
    \begin{split}
        G_{\bm{k}_1,\bm{k}_2}^{\bm{q}}&=F_{\bm{k}_1,\bm{k}_2}^{\bm{q}}\cdot \bigg[c_{\bm{k}_1}^*c_{Mod(\bm{k}_1-\bm{q})}c_{\bm{k}_2}^*c_{Mod(\bm{k}_2+\bm{q})}\\
        &+c_{\bm{k}_1+\bm{b}_2}^*c_{Mod(\bm{k}_1-\bm{q})+\bm{b}_2}c_{\bm{k}_2}^*c_{Mod(\bm{k}_2+\bm{q})}\cdot e^{\frac{i}{2}\bm{b}_2\times (Mod(\bm{k}_1-\bm{q})-\bm{k}_1)}\\
        & + c_{\bm{k}_1}^*c_{\bm{k}_4}c_{\bm{k}_2+\bm{b}_2}^*c_{Mod(\bm{k}_2+\bm{q})+\bm{b}_2}\cdot e^{\frac{i}{2}\bm{b}_2\times (Mod(\bm{k}_2+\bm{q})-\bm{k}_2)} \\
        &+c_{\bm{k}_1+\bm{b}_2}^*c_{Mod(\bm{k}_1-\bm{q})+\bm{b}_2}c_{\bm{k}_2+\bm{b}_2}^*c_{Mod(\bm{k}_2+\bm{q})+\bm{b}_2} \cdot e^{\frac{i}{2}\bm{b}_2\times (Mod(\bm{k}_1-\bm{q})-\bm{k}_1+Mod(\bm{k}_2+\bm{q})-\bm{k}_2)} \bigg]
    \end{split}
\end{equation}
Thus, the density-density interaction on the dLL can also be understood as an effective interaction projected onto the LLL, as shown in above equations.

\section{ Topological phases realized by screened Coulomb interaction at $\nu_{dLL}=1/3$}

\begin{figure}[ht]
\centering
\includegraphics[width=0.6 \linewidth]{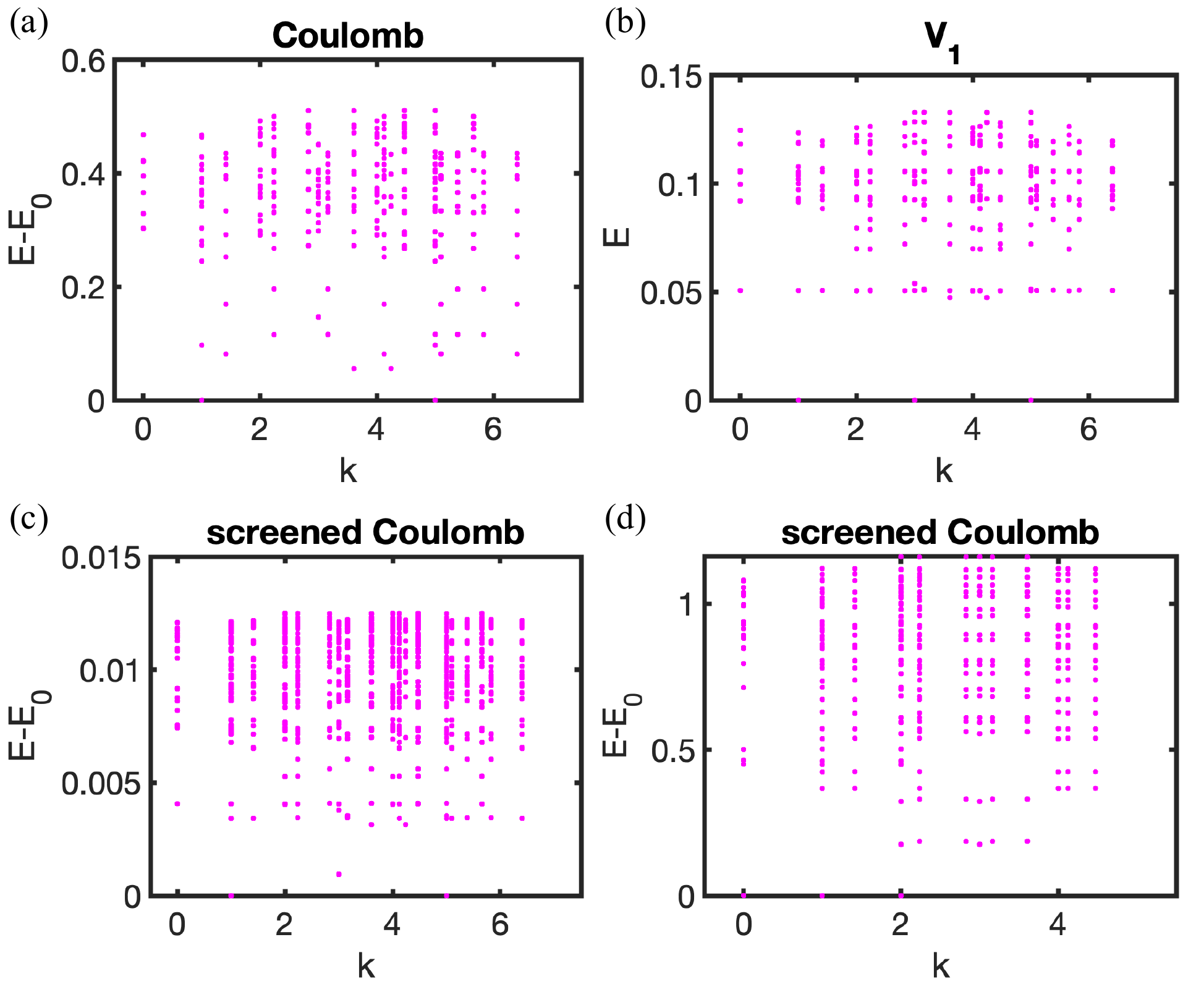}
\caption{The spectra of two-body interaction. (a) Coulomb interaction on the dLL at $\nu_{l}=1/6$. (b) Model Hamiltonian $V_1$ within the dLL at $\nu_{l}=1/6$. (c) Screened Coulomb interaction Yukawa with $d=0.4$ on the dLL at $\nu_{l}=1/6$. (d) Screened Coulomb interaction Yukawa with $d=1$ within the LLL Hilbert space at $\nu_{l}=2/3$.}
\label{figs1}
\end{figure}

Considering the whole LLL, when the dispersive bands are empty and dLL is one-third filled, model Hamiltonian $V_1$ governs the system, offering Laughlin states on the dLL. These states remain exact zero-energy states, with wavefunctions given in the previous section, and are precisely the Laughlin states in the LLL with a number of quasi-holes. However, the conformal Hilbert space is only conserved for the model Hamiltonian, and Coulomb interaction may not support the topological order on the dLL. The spectra of Coulomb interaction and $V_1$ projected onto the dLL are shown in fig.~\ref{figs1}(a) and (b), in which the number of total flux on the LLL $N_{\phi}=60$, $N_{\delta}=30$ and $N_e=10$. The vanishing quasi-degeneracy of ground states and FQH gap indicate that Coulomb interaction doesn't stabilize a topological phase. However, one can consider a short-range Yukawa interaction that is commonly used to incorporate the screening effect theoretically, 
\begin{equation}
\begin{split}
    & V(\bm{r})=\frac{e^{-|\bm{r}|/d}}{|\bm{r}|},\\
    & V(\bm{q})=\frac{1}{\sqrt{1/d^2+|\bm{q}|^2}},
\end{split}
\end{equation}
where $d$ is the inverse of coherent length in units of $\ell_B$. The spectra of Yukawa interaction on the dLL is shown in fig.~\ref{figs1}(c) at filling factor $\nu_l=1/6$. Even though the three quasi-degenerate ground states don't have the exactly same energy, the existence of FQH gap implies robustness of topological phases. This occurs because a larger coherence length leads to stronger screening of the Yukawa interaction, reducing the ratio $V_3/V_1$
. Consequently, the interaction becomes more 
like $V_1$, and the resulting spectrum approaches that of the pure $V_1$ case, where a gapped phase is realized. Thus, for a short-range Yukawa interaction, the gapped topological phase appears as shown in fig.~\ref{figs1}(c). Moreover, when the lattice potential dominates and $\lambda<0$ at $\nu_l=2/3$, which means the dispersive band is fully filled and the filling factor of the dLL is $1/3$, topological phase governed by the screened Coulomb interaction also appears on the dLL with strong mixing between the dLL and the dispersive band, as shown in the fig.~\ref{figs1}(d), in which $N_e=20$, $N_{\phi}=30$ and $N_{\delta}=15$.

\section{ Topological phases  at $\nu_{dLL}=1/5$}

\begin{figure}[ht]
\centering
\includegraphics[width=0.65 \linewidth]{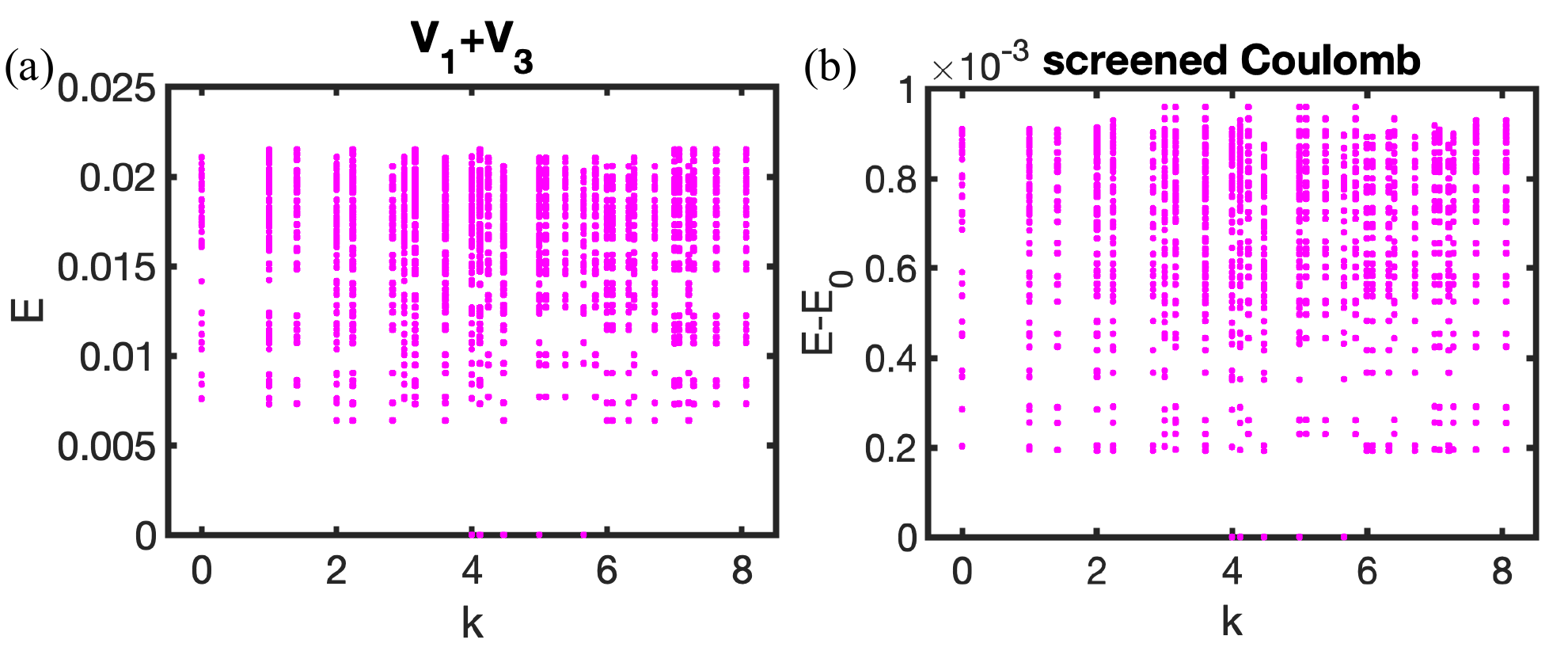}
\caption{The spectra of two-body interaction (a) model Hamiltonian $V_1+V_3$ on the dLL at $\nu_{l}=1/10$. (b) Screened Coulomb interaction Yukawa with $d=0.5$ within the dLL at $\nu_{l}=1/10$.}
\label{figs2}
\end{figure}

The conservation of conformal Hilbert space, not only remains for the Laughlin1/3 state, but also for Laughlin 1/5 phase. Laughlin 1/5 phase which spans the null space of model Hamiltonian $\alpha V_1+\beta V_3$ that $\alpha \neq 0$ and $\beta \neq 0$ is conserved on the dLL when such model Hamiltonian projected onto the dLL and the filling factor on the dLL $\nu_{dl}=1/5$, as mentioned in supplementary materials D. The spectra of $V_1+V_3$ on the dLL is shown in fig.~\ref{figs2}(a), the exact zero energy states with degeneracy indicates Laughlin 1/5 phase. Moreover, the screened Coulomb can also stabilize the Laughlin 1/5 state, as shown in fig.~\ref{figs2}(b), demonstrated by the five manifold ground states and stable FQH gap. Here we take $p=2$, $q=1$, $N_e$=8, $N_{\delta}=40$ and $N_{\phi}=80$. Laughlin 1/5 phase may occur when the screened Coulomb, Yukawa interaction, has a middle range such that $V_1$ is large enough but not completely suppress $V_3$, thus both $V_1$ and $V_3$ matter to introduce such a topological phase. Thus, in such case, the five quasi-degenerate states offer the conductivity $\sigma_{xy}=\frac{e^2}{5h^2}$ even though $\nu_{l}=1/10$, which is verified numerically.

\section{ Moore-Read phase at $\nu_{dLL}=1/2$}

\begin{figure}[ht]
\centering
\includegraphics[width=0.35 \linewidth]{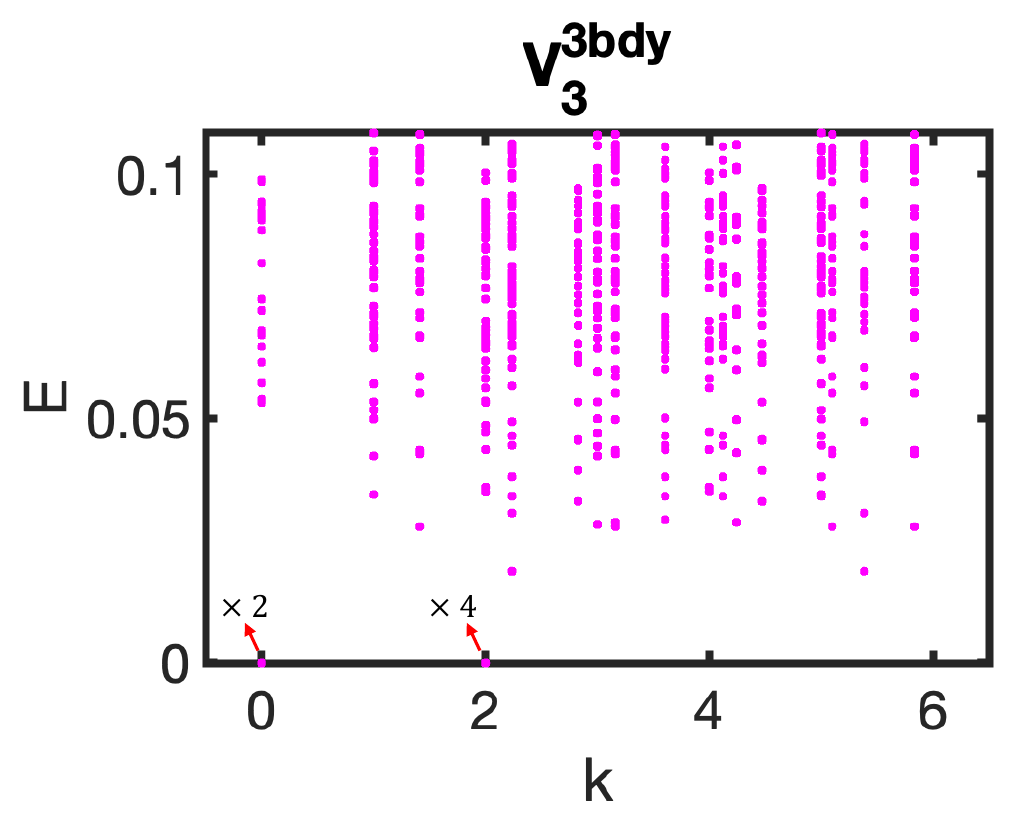}
\caption{The spectra of three-body interaction $V_3^{3bdy}$ at $\nu_{l}=1/4$ and $\nu_{dl}=1/2$.}
\label{figs3}
\end{figure}

Considering a three-body interaction $V_3^{3bdy}$, in real-space, it can be written in the way of $\delta$ functions:
\begin{equation}
    \hat{V}_3^{3bdy}=-\sum_{i<j<k} S_{i,j,k} \left[ \Delta_i^4 \Delta_j^2 \delta^2(\textbf{r}_i-\textbf{r}_j)\delta^2(\textbf{r}_j-\textbf{r}_k)\right],
\end{equation}
where $S_{i,j,k}$ is a symmetrizer going through the electrons $i,j,k$. The zero energy states of such Hamiltonian, span a Moore-Read conformal Hilbert subspace of the LLL. Similarly, the band projected three-body interaction can be written as,
\begin{equation}
    \hat{V}_3^{3bdy}=\sum_{\bm{k}_1,\bm{k}_2,\bm{k}_3,\bm{k}_4,\bm{k}_5,\bm{k}_6} V_{\bm{k}_1,\bm{k}_2,\bm{k}_3,\bm{k}_4,\bm{k}_5,\bm{k}_6}\delta'(\bm{k}_1+\bm{k}_2+\bm{k}_3,\bm{k}_4+\bm{k}_5+\bm{k}_6) c^{\dagger}_{\bm{k}_1} c^{\dagger}_{\bm{k}_2} c^{\dagger}_{\bm{k}_3} c_{\bm{k}_4} c_{\bm{k}_5} c_{\bm{k}_6},
\end{equation}
in which 
\begin{equation}
    V_{\bm{k}_1,\bm{k}_2,\bm{k}_3,\bm{k}_4,\bm{k}_5,\bm{k}_6}=\sum_{\bm{b},\bm{b}'}V(\bm{k}_1-\bm{k}_6-\bm{b},\bm{k}_2-\bm{k}_5-\bm{b}')f_{\bm{b}}^{\bm{k}_1,\bm{k}_6}f_{\bm{b}'}^{\bm{k}_2,\bm{k}_5}f_{\Delta \bm{b}-\bm{b}-\bm{b}'}^{\bm{k}_3,\bm{k}_4},
\end{equation}
in which 
\begin{equation}
    \Delta \bm{b}= (\bm{k}_1+\bm{k}_2+\bm{k}_3)-(\bm{k}_4+\bm{k}_5+\bm{k}_6),
\end{equation}
\begin{equation}
    V(\bm{q}_1,\bm{q}_2)=-|\bm{q}_1|^2\cdot |\bm{q}_1-\bm{q}_2|^4.
\end{equation}
Let's still take the case of $p=2$ and $q=1$, the projection of the delta three-body interaction onto the dLL can be written into,
\begin{equation}
        \hat{V}_3^{3bdy}=\sum_{\bm{k}_1,\bm{k}_2,\bm{k}_3,\bm{k}_4,\bm{k}_5,\bm{k}_6} U_{\bm{k}_1,\bm{k}_2,\bm{k}_3,\bm{k}_4,\bm{k}_5,\bm{k}_6}\delta'(\bm{k}_1+\bm{k}_2+\bm{k}_3,\bm{k}_4+\bm{k}_5+\bm{k}_6) c^{\dagger}_{\bm{k}_1} c^{\dagger}_{\bm{k}_2} c^{\dagger}_{\bm{k}_3} c_{\bm{k}_4} c_{\bm{k}_5} c_{\bm{k}_6}. 
\end{equation}
And we have,
\begin{equation}
\begin{split}
        U_{\bm{k}_1,\bm{k}_2,\bm{k}_3,\bm{k}_4,\bm{k}_5,\bm{k}_6} & = \sum_{\bm{b},\bm{b}'}V(\bm{k}_1-\bm{k}_6-\bm{b},\bm{k}_2-\bm{k}_5-\bm{b}') \cdot \\
        & \left[ c_{\bm{k}_1}^*c_{\bm{k}_6}f_{\bm{b}}^{\bm{k}_1,\bm{k}_6}+c_{\bm{k}_1+\bm{b}_2}^*c_{\bm{k}_6+\bm{b}_2}f_{\bm{b}}^{\bm{k}_1+\bm{b}_2,\bm{k}_6+\bm{b}_2}\right] \cdot \left[ c_{\bm{k}_2}^*c_{\bm{k}_5}f_{\bm{b}'}^{\bm{k}_2,\bm{k}_5}+c_{\bm{k}_2+\bm{b}_2}^*c_{\bm{k}_5+\bm{b}_2}f_{\bm{b}'}^{\bm{k}_2+\bm{b}_2,\bm{k}_5+\bm{b}_2}\right] \cdot \\
        & \left[c_{\bm{k}_3}^*c_{\bm{k}_4}f_{-\bm{b}-\bm{b}'+\Delta\bm{b}}^{\bm{k}_3,\bm{k}_4}+c_{\bm{k}_3+\bm{b}_2}^*c_{\bm{k}_4+\bm{b}_2}f_{-\bm{b}-\bm{b}'+\Delta\bm{b}}^{\bm{k}_3+\bm{b}_2,\bm{k}_4+\bm{b}_2}\right]. \\
\end{split}
\end{equation}

The spectra of the three-body interaction projected onto the dLL are given in fig.~\ref{figs3}, in which we take $N_e=12$, $N_{\phi}=48$, and $N_{\delta}=24$. There are two degenerate ground states in the crystal momentum sector $(0,0)$, and four degenerate ground states in the sector $(0,2)$. The 6-manifold zero energy states on the dLL indicate the conservation of the topological phase, leading to the conductivity $\sigma_{xy}=\frac{e^2}{2h}$ while $\nu_l=1/4$. And in such case, lattice potentials dominates and all electrons are trapped on the dLL, the Moore-Read gap implies the robustness of topological phase on the dLL.

\section{ Graviton modes in decorated Landau levels}
In this section, we provide additional details on the graviton modes (GMs) in the decorated Landau levels (dLLs), together with explicit numerical results from exact diagonalization and spectral function calculations that complement the discussion in the main text.

\begin{figure}[ht]
\centering
\includegraphics[width=\linewidth]{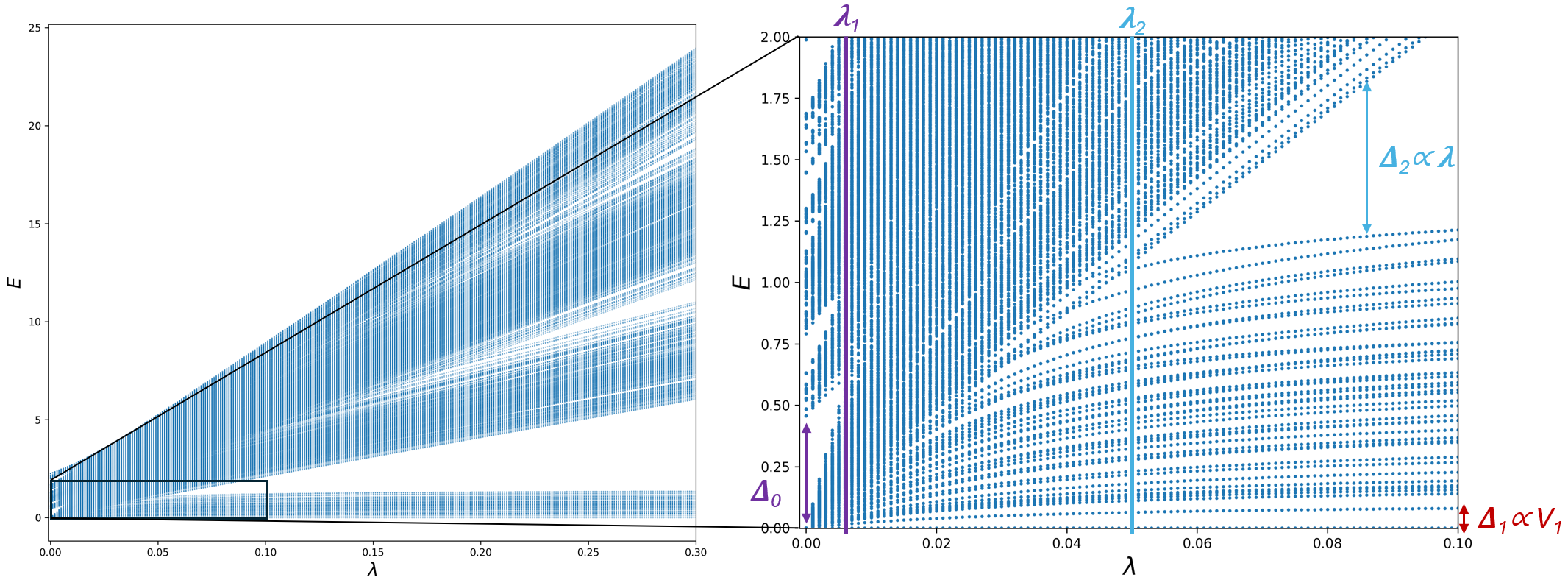}
\caption{
Spectrum of $\hat{H} = \hat{V}_1 + \lambda\, \hat{V}_\delta$ as a function of $\lambda$ for a system with $N_e = 4$ and $N_o = 24$ orbitals, with $N_\delta = 12$ delta potentials such that $\nu_{dl}=1/3$ in the dLL. The plot is taken within one of the ground-state momentum sectors. For $\lambda>0$ the ground state remains threefold degenerate, and the low-energy spectrum exhibits two characteristic $\lambda_1$ and $\lambda_2$ associated with the gap evolution discussed in the main text.}

\label{figj1}
\end{figure}

We begin by examining the evolution of the many-body spectrum of the Hamiltonian
\begin{equation}
    \hat{H} = \hat{V}_1 + \lambda\, \hat{V}_\delta ,
\end{equation}
where $\hat{V}_1$ is the Haldane pseudopotential, which serves as the model Hamiltonian for the Laughlin state at filling $\nu = 1/3$, and
\begin{equation}
    \hat{V}_\delta = \sum_{i=1}^{N_\delta} \delta(\hat{r} - R_i)
\end{equation}
is the lattice of delta potentials that generates the dLL structure. The full spectrum as a function of $\lambda$ is shown in Fig.~\ref{figj1} for a system with $N_e = 4$ electrons and $N_o = 24$ orbitals, with $N_\delta = 12$ delta sites chosen such that the effective filling within the dLL is $\nu_{dl} = 1/3$.

The left panel of Fig.~\ref{figj1} displays the complete spectrum within one of the three ground-state momentum sectors (the ground-state degeneracy remains three at zero energy for all $\lambda > 0$). The right panel zooms in on the low-energy region, where the two critical values $\lambda_1$ and $\lambda_2$ and the three characteristic gap scales discussed in the main text can be clearly identified. Increasing the system size does not qualitatively modify this gap-opening behavior, although the precise numerical values of $\lambda_1$ and $\lambda_2$ may shift due to finite-size effects.

We now briefly review the construction of graviton modes (GMs) in a Chern band. Within the conventional single–mode approximation (SMA), the GM is obtained by acting on the ground state with a regularized, Chern–band–projected density operator,
\begin{equation}
\left|\psi_{g}\right\rangle \sim  
\lim_{|\boldsymbol{q}|\rightarrow 0} 
\delta \hat{\bar{\rho}}_{\boldsymbol{q}} 
\left|\psi_0\right\rangle,
\end{equation}
which provides a universal route to access the neutral spin–$2$ collective excitation. However, on a finite torus, the limit $|\boldsymbol{q}|\to 0$ cannot be reached exactly, and the accuracy of the SMA is limited by finite–size momentum resolution.

A complementary and numerically more robust approach is provided by the two–body \textit{chiral graviton operator}, which probes the projected stress tensor \cite{GM2}:
\begin{equation}
\hat{O}_{\pm} = 
\sum_{\boldsymbol{q}} 
(q_x \pm i q_y)^2\, V(\boldsymbol{q})\,
\hat{\bar{\rho}}_{\boldsymbol{q}}
\hat{\bar{\rho}}_{-\boldsymbol{q}},
\label{eq:Opm}
\end{equation}
where $(q_x \pm i q_y)^2$ encodes the graviton chirality $\sigma$, and $V(\boldsymbol{q})$ is the interaction potential including the appropriate form factors. For the model Hamiltonian with $V(\boldsymbol{q}) = V_1(q)$, it can be shown analytically that the GM of the Laughlin state at $\nu = 1/3$ is perfectly chiral with $\sigma = -1$, a result that is also confirmed by our numerical calculations. All results presented below therefore focus on the $\sigma = -1$ graviton sector. The SMA operator and $\hat{O}_{\pm}$ are related through Ward identities, but as discussed above, in finite systems $\hat{O}_{\pm}$ more directly captures the interaction–driven metric fluctuations and is thus better suited for numerical implementation on the torus.

\begin{figure}[t]
\centering
\includegraphics[width=\linewidth]{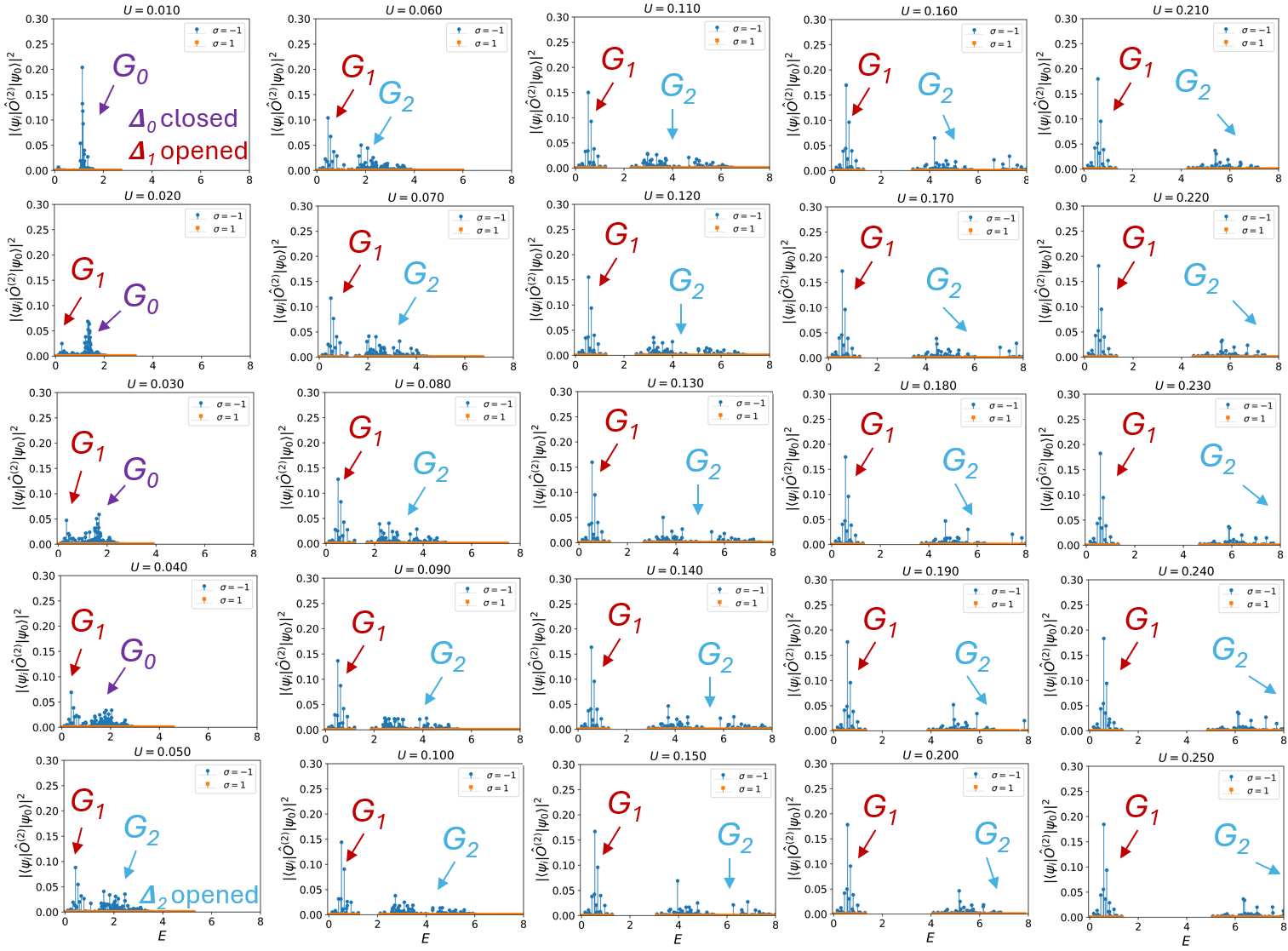}
\caption{
Graviton spectral function $I_{-}(E)$ for the chiral graviton operator $\hat{O}_{-}$ as a function of $\lambda$ for the same system size as in Fig.~\ref{figj1}. Each panel corresponds to a different value of $\lambda \in [0.01,0.5]$ (Here, $U$ is equivalent to the strength of the delta potentials $\lambda$), and the peak intensity indicates the spectral weight of the graviton excitation. For $\lambda \le \lambda_1$, the response is dominated by a single graviton peak $G_0$ associated with the conventional Laughlin graviton. In the intermediate regime $\lambda_1 < \lambda \le \lambda_2$, the spectral weight splits into two branches: a new dLL graviton peak $G_1$ that develops inside the dLL and a higher energy feature $G_2$ outside the dLL. For $\lambda > \lambda_2$, $G_1$ saturates as the dominant graviton mode within the dLL, while $G_2$ shifts to higher energies and its spectral weight is progressively suppressed, consistent with the finite size scaling analysis in the main text.
}
\label{figj2}
\end{figure}

Experimentally, signatures of GMs can be detected through circularly polarized Raman scattering, where the response is captured by the graviton spectral function
\begin{equation}
I_{\pm}(E) =
\sum_{n}
\big|\langle n | \hat{O}_{\pm} | 0 \rangle\big|^2\,
\delta(E - E_n + E_0),
\label{eq:spectral}
\end{equation}
with $\hat{O}_{\pm}$ defined in Eq.~\eqref{eq:Opm}. The position of the resonance peak determines the GM energy, while the linewidth reflects the GM lifetime. This provides a direct quantitative framework to compare the chiral GM behavior in decorated and conventional LLs, and to examine how the local potential lattice modifies their geometric response.

We now analyze how the spectral signatures of the GM evolve as the system is tuned from the interaction–dominated regime to the lattice–dominated regime by varying $\lambda$. In finite–sized systems, the spectral function exhibits a clear separation into two graviton peaks. Their evolution with $\lambda$ is shown in Fig.~\ref{figj2} for the same system size as in Fig.~\ref{figj1}. The behavior may be summarized as follows:

\textbf{(i) Small–$\lambda$ regime ($\lambda \le \lambda_1$):}
The spectral weight is concentrated in a single graviton peak $G_0$, the energy of which is approximately equal to the conventional LLL GM. The peak position lies deep within the excitation continuum, separated from the ground state by the interaction scale set by $V_1$, and its linewidth increases slightly as $\lambda$ increases.

\textbf{(ii) Intermediate regime ($\lambda_1 < \lambda \le \lambda_2$):}
The original graviton peak $G_0$ response splits into two components. A new peak $G_1$ develops within the dLL and gains spectral weight with increasing $\lambda$, while the residual weight of $G_0$ is transferred to a higher–energy feature $G_2$ associated with states outside the dLL. This regime marks the crossover between the Laughlin–like graviton and the intrinsic dLL graviton.

\textbf{(iii) Lattice–dominated regime ($\lambda > \lambda_2$):}
The $G_1$ peak saturates and forms the dominant graviton branch within the dLL, whereas $G_2$ is pushed to higher energies and its spectral weight is strongly suppressed with increasing system size. As discussed in the main text, this trend indicates that $G_2$ is likely a finite–size remnant that disappears in the thermodynamic limit, leaving a single dLL GM.

\begin{figure}[ht]
\centering
\includegraphics[width=\linewidth]{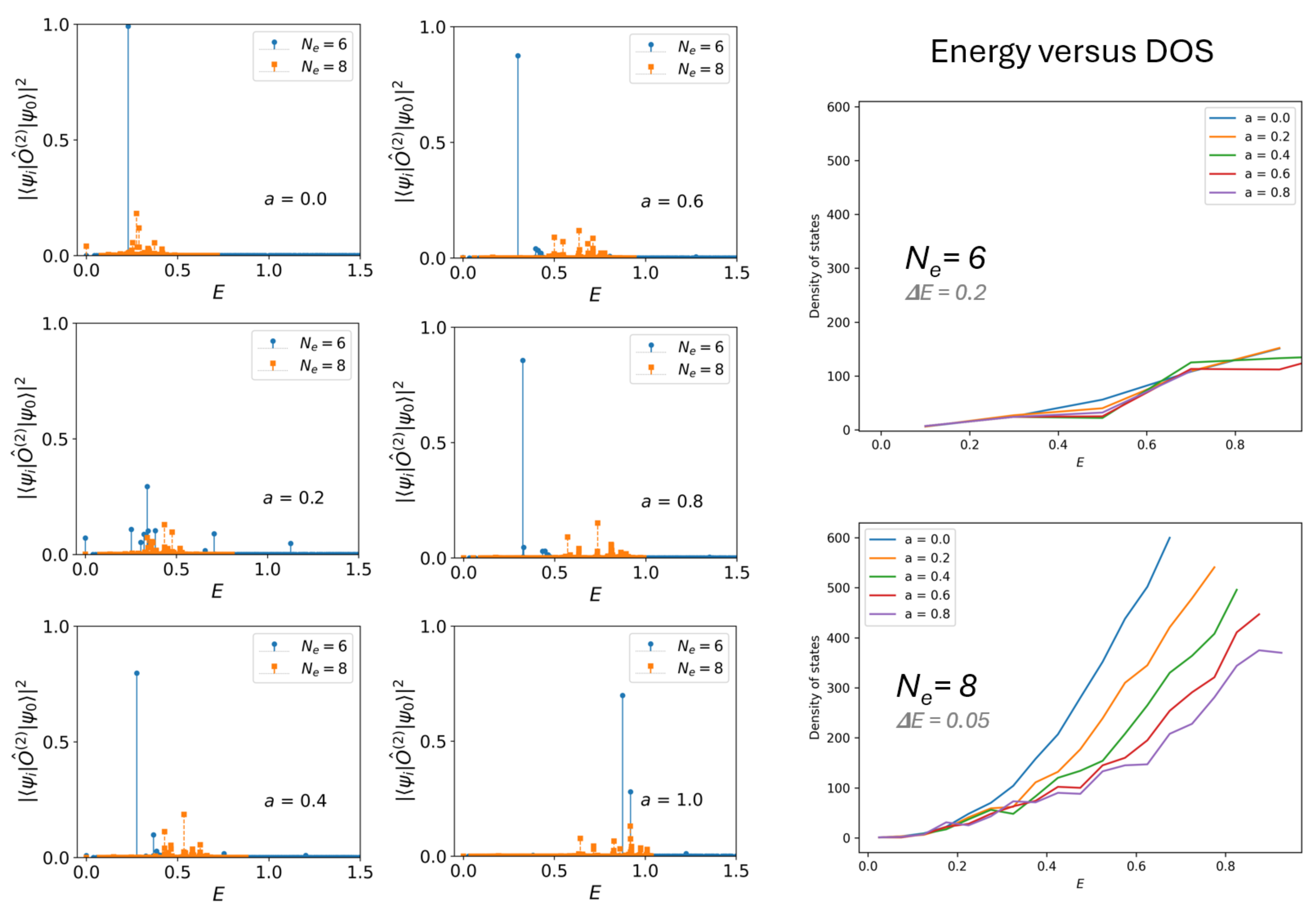}
\caption{
Spectral function of the dLL graviton $G_1$ in the lattice dominated regime, evaluated using the interpolating Hamiltonian $\hat{H}(a) = a\,\hat{V}_1 + (1-a)\,\hat{V}_{C}$, where $a$ tunes the position of the graviton relative to the excitation continuum. The left panels show the graviton spectral weight for system sizes $N_e = 6$ and $N_e = 8$ at different values of $a$. For $N_e = 6$, the graviton peak remains sharp for all $a$, whereas for $N_e = 8$ the peak becomes significantly broader, indicating a strong sensitivity to the surrounding continuum. The right panels display the corresponding density of states (DOS) near the graviton energy. The DOS is much lower in the $N_e = 6$ system than in the $N_e = 8$ system, which explains the artificially enhanced graviton peak in smaller system sizes. These results show that the apparent sharp graviton peak in small systems arises from finite-size spectral sparsity, and that the dLL graviton in the lattice-dominated regime is very likely to acquire a short lifetime in the thermodynamic limit.
}

\label{figj3}
\end{figure}

We now examine the lifetime of the dLL graviton mode. In finite systems, an apparent sharp peak in the spectral function does not necessarily imply a long–lived excitation in the thermodynamic limit, since the peak height may be artificially enhanced by a sparse local density of states (DOS). To get intrinsic lifetime effects without such finite–size artifacts, we introduce the toy Hamiltonian
\begin{equation}
\hat{H}(a) = a\,\hat{V}_1 + (1-a)\,\hat{V}_{C},
\end{equation}
where $a \in [0,1]$ continuously tunes the relative strength of the $V_1$ pseudopotential and the Coulomb interaction. Varying $a$ shifts the position of the graviton in the excitation continuum and thereby changes the DOS in its vicinity: increasing $a$ ``penalizes'' the Laughlin graviton more strongly and pushes it to a higher–energy region with a larger DOS.

This construction provides a controlled diagnostic of the graviton lifetime. If a graviton mode remains sharp and carries a large spectral weight for all values of $a$, its lifetime can be expected to remain long in the thermodynamic limit (as is the case for the Laughlin graviton in the LLL \cite{GM3}). In contrast, if the peak height is strongly dependent on $a$ or collapses when the DOS increases, then the sharp peak observed in small systems is likely a finite–size artifact rather than a robust long–lived collective mode.

We apply this analysis to the $G_1$ graviton in the lattice–dominated regime. The left panels of Fig.~\ref{figj3} show the spectral function of $\hat{O}_{-}$ for two system sizes, $N_e = 6$ and $N_e = 8$, as $a$ is varied. For $N_e = 6$, the graviton peak remains high and sharp across all $a$, seemingly suggesting an apparently stable excitation. However, when the system size is increased to $N_e = 8$, the same peak becomes significantly lower and more broadened, indicating a strong sensitivity to the nearby continuum and a reduced lifetime.

To confirm this interpretation, the right panels of Fig.~\ref{figj3} display the DOS around the graviton energy for the two system sizes. The $N_e = 6$ system exhibits a much lower DOS in the relevant energy window, explaining the artificially enhanced peak amplitude. In contrast, the $N_e = 8$ system has a substantially larger DOS, and the graviton peak correspondingly spreads out. Together, these results indicate that the dLL graviton in the lattice–dominated regime acquires a short lifetime, and that the sharp peaks observed in small systems arise predominantly from finite–size spectral sparsity rather than from a stable excitation. Taken together, these results suggest that the dLL graviton can only be probed experimentally if it is tuned out of the excitation continuum and brought into a spectral gap, for example, through external control of band geometry or interaction types. A systematic exploration of such tuning mechanisms is an interesting direction for future work.


\begin{thebibliography}{99}
\bibitem{footnote} We use the term ``decorated LL'' to distinguish it from the commonly used term ``modulated LL'', which usually refers to modifying Landau levels with a periodic magnetic field. Here, decorated LLs arise from modifying Landau levels with periodic electrostatic potentials. Notably, decorating LLs with delta-function potentials can realize exactly flat Chern bands, analogous to modulated LLs in graphene.
\bibitem{footnote3} Fig. ~\ref{fig5} (c) and (d) are fitted with some exact diagonalization data points, thus may omit some phases due to strong finite size effect and odd-even effect. When $\lambda<0$, $\nu_{dl}=1/3$ and dispersive bands are fully filled, we conjecture topological phases may occur even though $p/q$ is not well designed for larger systems.
\end{thebibliography}
\end{document}